\pdfoutput=1
\documentclass[usenatbib,usedAMS]{mnras}

\usepackage{natbib}
\usepackage{amsmath}
\usepackage{graphicx}
\usepackage{color}
\usepackage{mathrsfs}
\usepackage{epsfig}
\usepackage{comment}
\usepackage{ulem}

\newcommand{\msun}{{\rm M}_\odot}

\newcommand{\gcc}{{\rm g~cm}^{-3}}
\newcommand{\cc}{{\rm cm}^{-3}}

\newcommand{\msunyr}{{\rm M}_\odot~{\rm yr}^{-1}}

\newcommand{\kpc}{{\rm kpc}}

\newcommand{\pc}{{\rm pc}}
\newcommand{\kms}{{\rm km~s}^{-1}}

\newcommand{\K}{{\rm K}}
\newcommand{\beq}{\begin{equation}}
\newcommand{\eeq}{\end{equation}}

\defcitealias{K14}{K14}
\defcitealias{K16}{K16}
\defcitealias{Visbal_2015}{VHB15}

\title[]
{
Transition of BH feeding from the quiescent regime into star-forming cold disk regime
}
\author[]{Kohei Inayoshi$^{1}$\thanks{E-mail: inayoshi@pku.edu.cn (KI)},
{Kohei Ichikawa}$^{2}$, {Jeremiah P. Ostriker}$^3$, 
and Rolf Kuiper$^4$
\\
\\
$^1$ Kavli Institute for Astronomy and Astrophysics, Peking University, Beijing 100871, China,\\
$^2$ Frontier Research Institute for Interdisciplinary Sciences, Tohoku University, Sendai 980-8578, Japan\\
$^3$ Department of Astronomy, Columbia University, 550 W. 120th Street, New York, NY 10027, USA,\\
$^4$ Institute of Astronomy and Astrophysics, University of T\"{u}bingen, Auf der Morgenstelle 10, D-72076 T\"{u}bingen, Germany}

\begin{document}
\maketitle

\begin{abstract}
We study the properties of rotating accretion flows onto supermassive black holes (SMBHs) using
axisymmetric two-dimensional hydrodynamical simulations with radiative cooling and BH feedback.
The simulations resolve the accretion dynamics of gas outside from the BH influence radius through an inner accretion disk.
For lower Bondi accretion rates in units of the Eddington rate ($\dot{M}_{\rm B}\ll 10^{-3}~\dot{M}_{\rm Edd}$),
the BH feeding is suppressed due to turbulent motion by several orders of magnitude from the Bondi rate
with outflows to the Bondi radius nearly balancing inflows.
Thus, the radiative luminosity results in as low as $\sim 10^{-10}-10^{-7}~L_{\rm Edd}$, 
where $L_{\rm Edd}$ is the Eddington luminosity.
For higher rates of $\dot{M}_{\rm B}\ga 10^{-3}~\dot{M}_{\rm Edd}$, the optically-thin accreting gas cools via free-free emission
and forms a geometrically-thin disk,
which feeds the BH efficiently and increases the radiative luminosity to $\ga 10^{-3}~L_{\rm Edd}$.
The transitional behavior of accreting BHs in galactic nuclei from radiatively inefficient phases to cold disk accretion naturally explains
(1) the reason for the offset between the observed luminosities and theoretical predictions for nearby quiescent SMBHs, and 
(2) the conditions to fuel gas into the nuclear SMBH.
In addition, the cold disk formed in galactic nuclei tends to be gravitationally unstable and leads to star formation
when the Bondi rate is as high as $ \dot{M}_{\rm B} \ga 10^{-2}~\msunyr$.
This is a plausible explanation of the correlation observed between star formation rates and BH feeding rates in Seyfert galaxies.
\end{abstract}

\section{Introduction}
\label{sec:intro}

Supermassive black holes (SMBHs) are almost ubiquitously harbored at the centers of massive nearby galaxies.
The existence of SMBHs is consistent with the number and energetics of high-redshift bright quasars (QSOs),
which are associated with efficient gas accretion onto SMBHs \citep{Soltan_1982,YT_2002}.
Through BH feeding and feedback processes, SMBHs are believed to coevolve with their host galaxies over the cosmic time
\cite[e.g.,][and references therein]{Kormendy_Ho_2013}. 
In the local universe, however only a few percent of SMBHs are observed as luminous active galactic nuclei (AGN).
A majority of them are nearly quiescent and known as low luminosity AGN 
\citep[][]{Ho_2008,Ho_2009}, or they may be undetectably faint.

While low-luminosity AGN are energetically unimpressive, the study of such AGN phenomena brings us important insights and 
intriguing questions regarding the physics of BH accretion.
One of the puzzles is that nearby, silent SMBHs exhibit levels of activities much lower than that 
expected from gas supplying rates onto the galactic nuclei \citep[e.g.,][]{Pellegrini_2005,Ho_2008,Ho_2009}.
This luminosity deficit problem clearly suggests the lack of our understanding on how accreting gas finally reaches to the central SMBHs.
Simultaneously, we need to address how such quiescent SMBHs turn into active phases, 
and how their host galaxies control the cycle of the two phases.

The basic physics of the accretion process was first studied for a spherically symmetric flow onto a BH with a mass of $M_\bullet$
without radiative effects \citep{Bondi_Hoyle_1944,Bondi_1952}.
This approach remains a good approximation only if rotation is so minor as to allow quasi-spherical accretion 
all the way down to the BH from the so-called Bondi radius defined by 
\begin{equation}
R_{\rm B}\equiv \frac{GM_\bullet}{c_s^2},
\end{equation}
within which the negative gravitational energy dominates the thermal energy of the gas with a sound speed of $c_s$,
where $G$ is the gravitational constant.
However, it is hard to conceive of cases where the gas could have such a tiny amount of angular momentum. 
The characteristic radius where the centrifugal force balances the gravity of the BH is given by
\begin{equation}
R_{\rm C}\equiv \frac{j^2}{GM_\bullet},
\end{equation}
where $j$ is the angular momentum per unit mass.
This radius may or may not be larger than the Bondi radius, but is always larger than the Schwarzschild radius,
\begin{equation}
R_{\rm Sch}\equiv \frac{2GM_\bullet}{c^2} \ll R_{\rm C}, R_{\rm B},
\end{equation}
where $c$ is the speed of light.
This means that the inflowing gas must always fall to a rotating (thin or thick) disk before reaching the BH,
and subsequently its mass transport inwards is determined by the physical processes that transport the 
angular momentum outwards \citep{Lynden-Bell_Pringle_1974,Pringle_1981}.

At sufficiently low accretion rates, the infalling matter is not dense enough to cool.
Thus, the flow is essentially adiabatic, and then there would be no net accretion in the absence of viscosity. 
However, magneto-hydrodynamical (MHD) simulations have shown that even small entrained magnetic 
fields will induce the magneto-rotational instability (MRI) in a sufficiently ionized disk 
\citep{Balbus_Hawley_1991,Matsumoto_1995,Stone_1996,Balbus_Hawley_1998,Stone_Pringle_2001,
Hawley_2001,Machida_2001,McKinney_&_Gammie_2004, Ohsuga_2009,Bai_2011,Narayan_2012,Suzuki_Inutsuka_2014}.
MRI-driven turbulence causes an effective shear viscosity which permits further inflows
and is often described with the $\alpha$-viscosity prescription \citep[e.g.,][]{SS_1973}.
In this very low accretion rate domain, several papers have shown that rotating accretion flows 
become convectively unstable and have been termed convection dominated accretion flows
\citep[CDAFs;][]{IA_1999,Stone_1999,Narayan_2000,Quataert_2000,IA_2000}.

\begin{figure}
\begin{center}
\includegraphics[width=83mm]{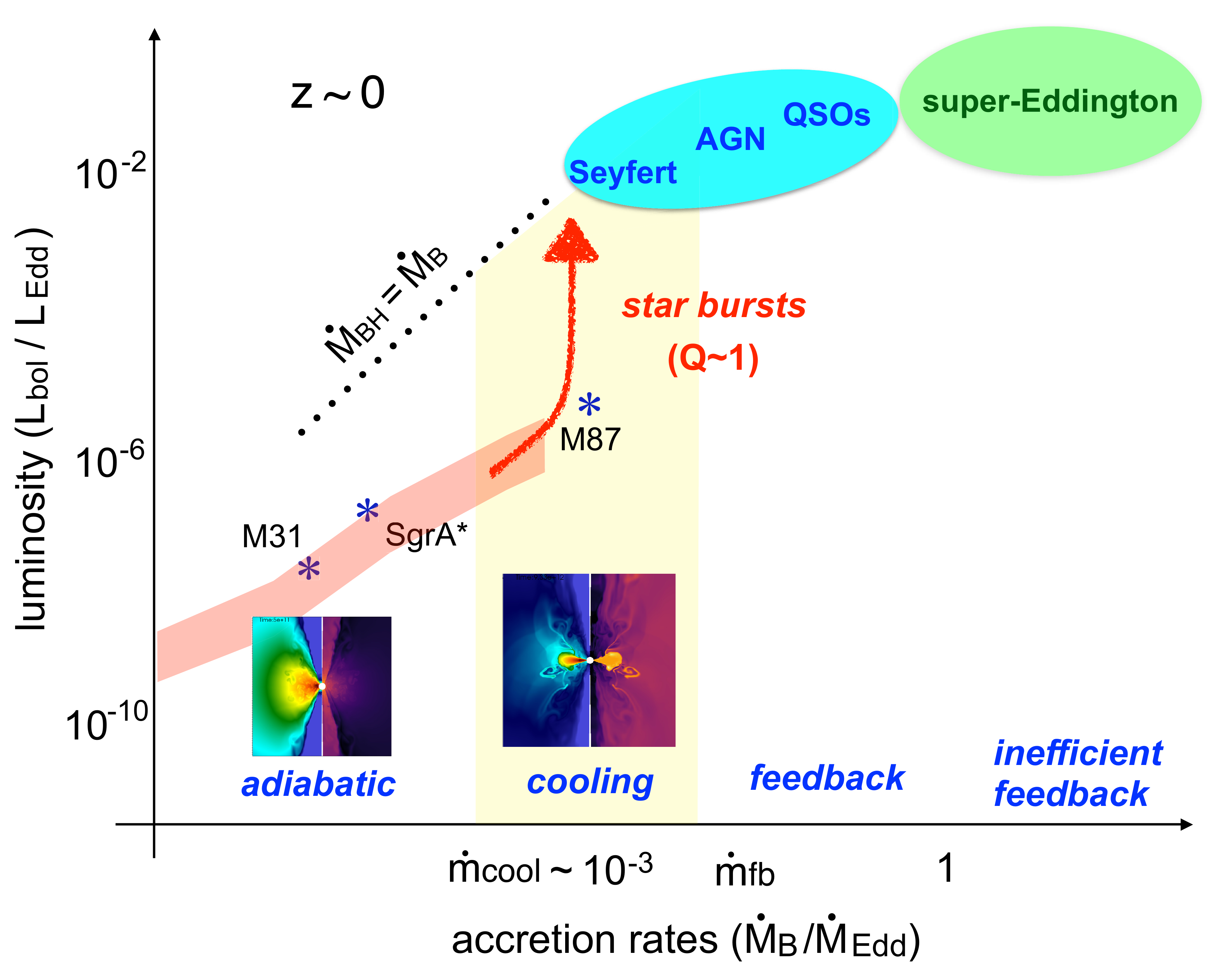}
\vspace{0mm}
\caption{A schematic overview of BH accretion solutions.
The characteristic bolometric luminosities are shown as a function of the gas inflow rates (i.e., the Bondi accretion rates).
Both values are normalized by the Eddington luminosity and accretion rate.
For lower rates at $\dot{M}_{\rm B}/\dot{M}_{\rm Edd}<10^{-3}(\sim \dot{m}_{\rm cool})$, 
the accreting gas behaves adiabatically and forms a geometrically-thick disk.
As a result, matter is convectively unstable and the BH accretion rate is well below the Bondi rate ($\dot{M}_\bullet \ll \dot{M}_{\rm B}$), 
and the radiative luminosity is as small as $L_{\rm bol}/L_{\rm Edd}\la 10^{-6}$ \citep{Inayoshi_2018}.
The pink band encloses the CDAF solutions which explain nearby silent SMBHs.
For higher $\dot{M}_{\rm B}/\dot{M}_{\rm Edd}\ga \dot{m}_{\rm cool}$, the gas cools and forms a geometrically-thin disk.
The BH accretion rate increases by several orders of magnitude and reaches the Bondi rate ($\dot{M}_\bullet \simeq \dot{M}_{\rm B}$),
which causes a sharp rise of the radiative luminosity.
The cold accretion disk is likely to be unstable by its self-gravity ($Q\sim 1$), 
fragment into small clumps and triggers star bursts.
}
\vspace{0mm}
\label{fig:diagram}
\end{center}
\end{figure}

Although other solutions of hot accretion flows have been proposed by many authors for different boundary conditions
\citep[e.g.,][]{Ichimaru_1977,NY_1994,NY_1995,Blandford_Begelman_1999,Blandford_Begelman_2004},
accreting matter captured by the BH gravity through the Bondi radius preferentially settles down to a CDAF within the centrifugal radius 
(\citealt{Inayoshi_2018}, Paper I).
In the CDAF solution, convective flows form a geometrically-thick disk, and the inflow rate decreases 
toward the center as $\dot{M}_{\rm in}\propto r$ within the disk. 
Physically, the reason for this is that the flow is not strongly bound to the central BH in the absence of cooling,
so that at each radius gravitational energy release is sufficient to reverse the flow of some material and consequently
to reduce the net accretion rate.
As a result, the net accretion rate onto the central BH is extremely small compared to the rate measured at the Bondi radius. 
As shown in a schematic overview in Fig. \ref{fig:diagram}, the solution describes the situation in the majority 
of the locally observed BHs such as the ones in the Milky Way, 
M31 up to the giant one in M87 \citep{Inayoshi_2018}, illustrated by the pink band in Fig. \ref{fig:diagram}.

Now, let us consider higher density environments which could be occasioned either by cooling flows 
within galaxies \citep[e.g.,][]{Ciotti_Ostriker_2001,Dekel_2006,Gaspari_2013,Gaspari_2015,Gan_2018} or galaxy mergers 
\citep[e.g.,][]{Sanders_1988,Juneau_2009,Hopkins_Quataert_2010}.
At some critical density, the rotating disk cools and collapses to a thin disc and then, as we shall see, 
in most cases the disks are locally unstable to the gravitational instabilities \citep{Toomre_1964,Gammie_2001}. 
Then spiral modes develop and provide the angular momentum and mass transport 
\citep[e.g.,][]{TQM_2005, Rice_2005, Lodato_Natarajan_2006, Hopkins_Quataert_2010, Kuiper_2011, Meyer_2017,Gan_2018}.
There is both observational and theoretical evidence for massive star formation to occur within the gravitationally unstable disks
\citep{Goodman_2003,Levin_Beloborodov_2003,Levin_2007,Nayakshin_2007}.

In this paper, we aim to pinpoint just where and when accretion onto BHs can change drastically 
from almost being unobservably faint to emitting more than a billion times the solar luminosity.
For this purpose, we perform axisymmetric two-dimensional hydrodynamic simulations including a variety of physical processes,
e.g., radiative cooling and feedback associated with BH accretion.
We investigate the gas dynamics of accretion in the cooling transition and elucidate the characteristics 
of the flow in the gravitationally unstable domain.
It occurs at the level where the accretion rate from the Bondi radius is $\dot{m}_{\rm cool}\sim O(10^{-3})$ 
measured in units of the Eddington accretion rate (see the yellow shaded region in Fig. \ref{fig:diagram}). 
There is another flow boundary at still higher inflow rates of $\dot{m}_{\rm fb} \sim O(10^{-2})$, 
where the radiative feedback from the central BH, i.e., gas heating at larger scales around $R_{\rm B}$, causes the flow to 
reverse and become intermittent.
This critical accretion rate has also been investigated by many authors
\citep{Ciotti_Ostriker_2001,Proga_2007,Ciotti_2009,Milosavljevic_2009,PR_2011,Novak_2011,
Choi_2014,IHO_2016} that basically focused on the feedback effect on the gas dynamics at $r\la R_{\rm B}$
or even larger galactic scales $\sim O(\kpc)$.
At even higher accretion rates of $\dot{m}\ga 1$, emergent photons are efficiently trapped within optically thick 
accretion flows \cite[e.g.,][]{Ohsuga_2005,Jiang_2014,Sadowski_2015}, and thus radiative feedback hardly affects 
gas supplied from the Bondi scales when the rate is as high as $\dot{m}\sim O(10^3)$ \citep{IHO_2016,Takeo_2018,Takeo_2019}.
We do not reach those levels of accretion in this work, but will study the feedback self-consistently 
coupled with the dynamics of the accretion disk in future work.

We here focus on the accretion dynamics at larger scales covering the Bondi radii, unlike previous work that has investigated
the accretion flows at smaller scales assuming a compact torus in hydrostatic equilibrium as the initial state.
This basic spirit of our simulations allows us to consider plausible initial and boundary conditions, some of which can be directly measured in the nuclear regions 
surrounding quiescent SMBHs.
As shown in Fig. \ref{fig:diagram}, we are also able to test our theoretical model compared to the energy output (e.g., radiative luminosity) 
associated with actual BH feeding at the centers of nearby galaxies.

The rest of this paper is organized as follows. 
In \S\ref{sec:method}, we describe the methodology of our numerical simulations. 
In \S\ref{sec:result}, we show our simulation results and explain their physical properties.
In \S\ref{sec:vis_cool}, we discuss the conditions for the transition from the radiatively inefficient, hot, torus-like accretion domain
to radiatively efficient, cold, thin disk accretion domain.
In \S\ref{sec:QQQ}, we discuss the possibility of gravitational instability of the disk and the properties of marginally unstable disk structure.
In \S\ref{sec:mdot-lumbol}, we describe the relation between the Bondi accretion rates and radiative luminosities and
explain the reason for the luminosity deficit of SMBHs in the local Universe.
Moreover, in \S\ref{sec:sfr} we argue the AGN-starburst connection, i.e., the relation between BH feeding rates and star formation rates 
in the circumnuclear regions.
Finally, we summarize our main conclusions in \S\ref{sec:sum}.


\section{Methodology}
\label{sec:method}

\subsection{Basic equations}

We solve the axisymmetric two-dimensional hydrodynamical equations using the open source code 
{\tt PLUTO} \citep{Mignone_2007} with modifications described in \cite{Kuiper_2010,Kuiper_2011}.
The basic equations are following: the equation of continuity,
\begin{equation}
\frac{d\rho}{dt}+\rho \nabla \cdot \mbox{\boldmath $v$}=0,
\end{equation}
and the equation of motion,
\begin{equation}
\rho \frac{d\mbox{\boldmath $v$}}{dt}=-\nabla p -\rho \nabla \Phi 
+ \nabla \cdot \mbox{\boldmath $\sigma$}
+f_{\rm rad}\mbox{\boldmath $e_r$},
\end{equation}
where $\rho$ is the density, {\boldmath $v$} is the velocity, and 
$p$ is the gas pressure,
the gravitational potential is set to $\Phi =-GM_\bullet /r$, 
$r$ is the distance from the central BH,
{\boldmath $\sigma$} is the stress tensor due to viscosity,
and $f_{\rm rad}$ is the outward net radiation force in the radial direction.
The time derivative is the Lagrangian derivative, given by 
$d/dt\equiv \partial/\partial t$ + {\boldmath $v$}$ \cdot \nabla$.

We solve the energy equation of
\begin{equation}
\rho \frac{de}{dt}=-p\nabla \cdot \mbox{\boldmath $v$} 
+ (\mbox{\boldmath $\sigma$} \cdot \nabla) \mbox{\boldmath $v$}
-\Lambda + \Gamma,
\end{equation}
where $e$ is the internal energy per mass.
The equation of state of the ideal gas is assumed as $p=(\gamma-1)\rho e$, 
where $\gamma = 5/3$.
The first and second terms on the right-hand-side present compressional heating 
(or expansion cooling) and viscous heating.
The last two terms are radiative cooling and heating (in units of ${\rm erg~s^{-1}~cm^{-3}}$).

The viscous stress tensor is given by
\begin{equation}
\sigma_{ij}=\rho \nu \left[ 
\left(\frac{\partial v_j}{\partial x_i} + \frac{\partial v_i}{\partial x_j}\right)
-\frac{2}{3} (\nabla \cdot \mbox{\boldmath $v$} )\delta_{ij}
\right],
\end{equation}
where $\nu$ is the shear viscosity.
Note that the bulk viscosity is neglected here.
The shear viscosity is calculated with the $\alpha$-prescription \citep{SS_1973},
\begin{equation}
\nu = \alpha \frac{c_{\rm s}^2}{\Omega_{\rm K}},
\label{eq:alpha_visc}
\end{equation}
where 
$\Omega_{\rm K}\equiv (GM_\bullet/r^3)^{1/2}$.
We calculate the viscous parameter by mimicking some properties of the MRI as
\begin{equation}
\alpha = \alpha_0 \left\{ \exp{\left[ 
-\left(\frac{\rho_{\rm cr}}{\rho}\right)^2\right]} + {\rm max}\left(0, -\frac{\partial \ln j}{\partial \ln r}\right)\right\},
\label{eq:alpha_j}
\end{equation}
where the strength of viscosity is set to $\alpha_0 =0.01$ according to MHD simulations of the global disk structure
\citep[e.g.,][]{Zhu_Stone_2018,Takasao_2018}, and 
$\rho_{\rm cr}$ is a threshold of the density above which viscosity turns on.
We adopt a maximum value of the density at an initial condition as the threshold value (see \S\ref{sec:ic}).
Under this model, the viscous process is active primarily in an accretion disk ($r\la R_{\rm C}$),
where the rotational velocity has a significant fraction of the Keplerian velocity.
On the other hand, angular momentum transported from the disk would be accumulated outside it, 
where no viscous processes operate.
Such rotating flows with negative gradients of the specific angular momentum outward, i.e., $\partial j/\partial r <0$, are unstable and 
become turbulent, leading to angular momentum transport in three-dimensional (3D) simulations \citep{Chandrasekhar_1961}.
Since our simulations do not capture this 3D effect, instead we add the second term in Eq. (\ref{eq:alpha_j})
so that a steady state of the accretion flow exists.
We note that the treatment of the rotational instability does not affect our results (see Appendix in \citealt{Inayoshi_2018}).

In our simulations, radiative processes are taken into account in what follows.
The radiative heating and cooling terms are expressed as
\begin{equation}
\Gamma - \Lambda = n^2(S_{\rm br} + S_{\rm Comp} + S_{\rm ph} ),
\label{eq:cool}
\end{equation}
where each term corresponds to the rate associated with 
bremsstrahlung cooling ($S_{\rm br}$), Compton heating/cooling ($S_{\rm Comp}$), and
the sum of photoionization heating, line and recombination continuum cooling ($S_{\rm ph}$).
The Compton heating/cooling rate is given by 
\begin{equation}
S_{\rm Comp} =\frac{k_{\rm B}(T_{\rm C}-T)}{m_{\rm e}c^2}
\frac{\sigma_{\rm es} L_{\rm bol}}{\pi n r^2},
\end{equation}
where $L_{\rm bol}$ is the AGN bolometric luminosity, $T_{\rm C}$ is the Compton temperature, 
$n$ is the number density of electron, $k_{\rm B}$ is the Boltzmann constant, $m_{\rm e}$ is the electron mass,
$c$ is the speed of light, and $\sigma_{\rm es}$ is the cross section of electron scattering.
We adopt $T_{\rm C}=10^8~\K$ as our fiducial value for observed low-luminosity AGN \citep*{Xie_2017}.
Note that Compton temperature for low-luminosity AGN ($L_{\rm X}/L_{\rm Edd}<10^{-3}$) 
is higher than that estimated in previous works using composite spectra of high luminosity AGNs; 
$T_{\rm C}\simeq 2\times 10^7~\K$ \citep{Sazonov_2004}.
The functional forms of $S_{\rm br}$ and $S_{\rm ph}$ are given in \cite{Sazonov_2005}.
In this paper, we consider situations where accretion flows are optically thin.
Therefore, the outward radiation force term in equation of motion can be expressed by 
\begin{equation}
f_{\rm rad}=\frac{\rho \kappa_{\rm es}}{c}\frac{L_{\rm bol}}{4\pi r^2} + \frac{\Gamma}{c}.
\end{equation}

\begin{figure}
\begin{center}
\includegraphics[width=80mm]{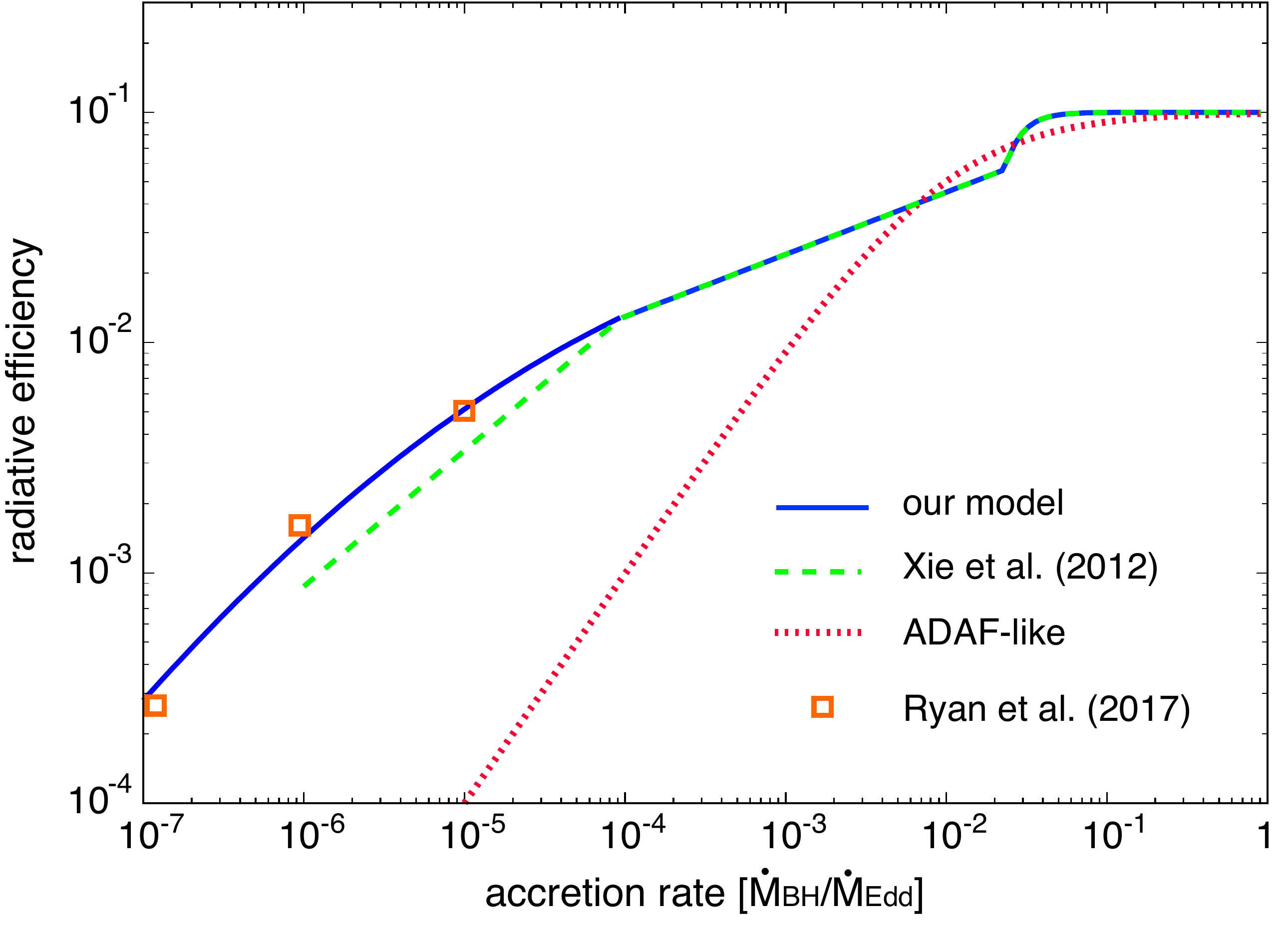}
\vspace{0mm}
\caption{
Radiative efficiency models for a hot accretion disk as a function of the BH feeding rate
in units of $\dot{M}_{\rm Edd}$.
Dashed (green) curve and symbols (orange square) show the efficiency obtained from a semi-analytical model 
\citep{Xie_Yuan_2012}, and by general relativity and radiation transfer simulations \citep{Ryan_2017}.
Solid (blue) curve presents the values we adopt in this work (linear interpolation of the log values at 
$\dot{M}_{\rm BH}/\dot{M}_{\rm Edd}<10^{-4}$).
Dotted (red) curve is an ADAF-like model \citep[e.g.,][]{Ciotti_2009}, which is often used in previous work.}
\label{fig:rad_eff}
\end{center}
\end{figure}

We estimate the bolometric AGN luminosity as $L_{\rm bol}=\epsilon \dot{M}_\bullet c^2$,
where $\dot{M}_\bullet$ is the gas accretion rate through the inner-most grid and 
$\epsilon$ is the radiative efficiency.
\begin{equation}
\log \epsilon = \begin{cases}
-1.0 -(0.0162/\dot{m})^4 & {\rm for}~~0.023\leq \dot{m},
\vspace{1mm}\\
\sum_{n} a_n (\log{\dot{m}})^n & {\rm for}~~~10^{-4}<\dot{m}<0.023,
\vspace{1mm}\\
\sum_{n} b_n (\log{\dot{m}})^n & {\rm for}~~~10^{-8}<\dot{m}\leq 10^{-4},
\end{cases}
\end{equation}
where $\dot{m}\equiv \dot{M}_\bullet/\dot{M}_{\rm Edd}$.
The fitted values are $a_0=-0.807$, $a_1=0.27$, $a_n=0$ ($n\geq 2$), $b_0=-1.749$, $b_1=-0.267$, $b_2=-0.07492$ and $b_n=0$ ($n\geq3$).
In Fig. \ref{fig:rad_eff}, we show the radiative efficiency model we adopt (blue solid curve),
which consists of the following two models.
The radiative efficiency at a higher rate of $\dot{m}>10^{-4}$ is obtained by a semi-analytical model \citep[green dashed curve;][]{Xie_Yuan_2012}.
For lower rates of $\dot{m}\leq 10^{-4}$, the fitted values are based on the result of general relativity and radiation transfer simulations 
(orange squares; see Table 1 in \citealt{Ryan_2017}), where gas accretion onto a Kerr BH with a spin of $a_\bullet=0.5$ has been studied.
Our model for the radiative efficiency is significantly higher than that for an ADAF-like model \citep[e.g.,][red dotted]{Ciotti_2009}, 
which is often used in previous work, but does not take into account energy transfer by Coulomb collision 
between ions and electrons properly.

\subsection{Boundary and initial conditions}
\label{sec:ic}

\begin{figure}
\begin{center}
\includegraphics[width=80mm]{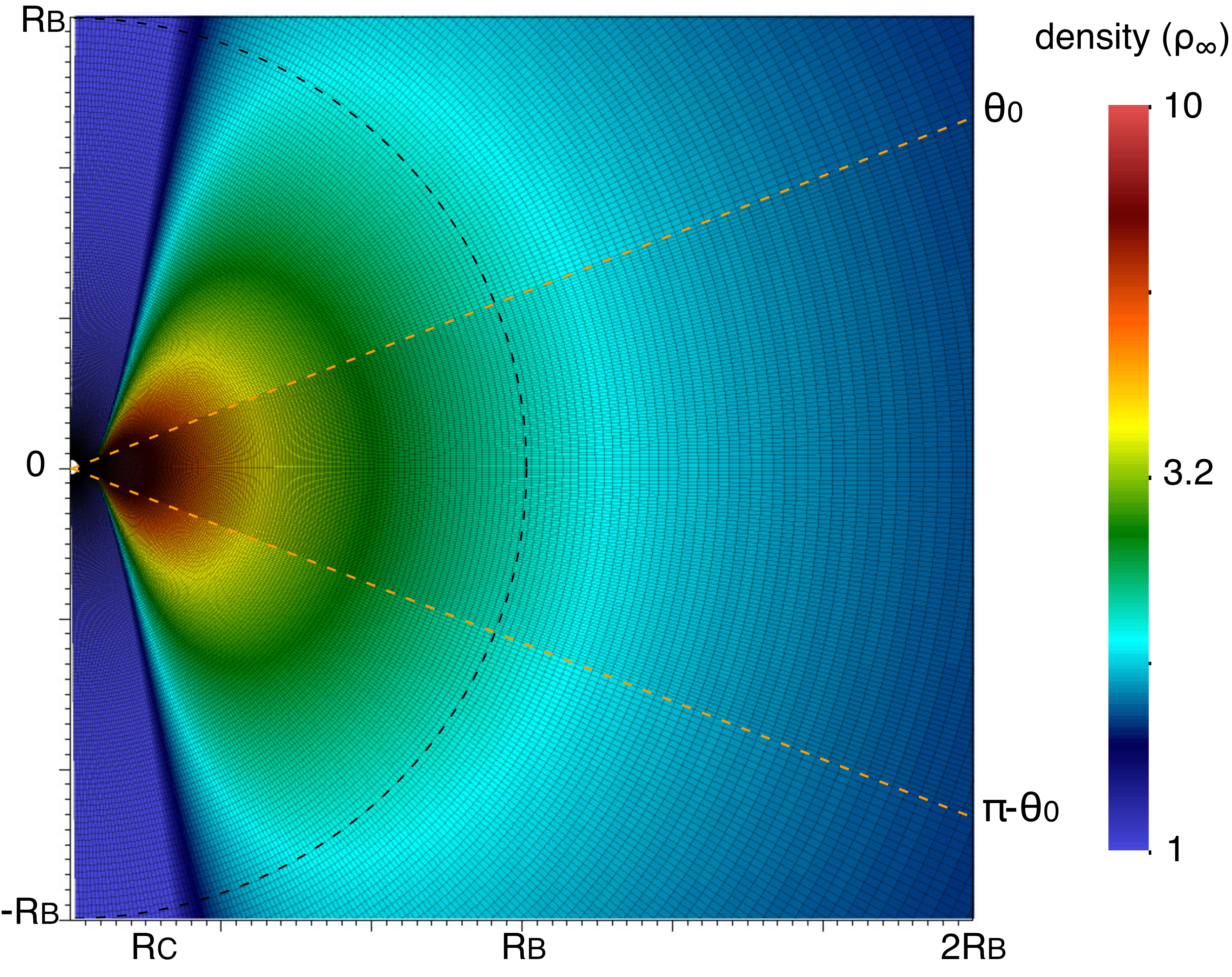}
\vspace{2mm}
\caption{Density distribution of the initial equilibrium torus given by Eq. (\ref{eq:torus}) for $R_{\rm C}/R_{\rm B}=0.1$.
The grid structure for our simulations is shown.
In the polar direction, we use static refinement at $\theta_0 \leq \theta \leq \pi-\theta_0$ (near the mid-plane)
in order to resolve the scale height of a cold disk (see \S\ref{sec:result}).
}
\vspace{-0mm}
\label{fig:ini}
\end{center}
\end{figure}

To compute the basic equations, we employ spherical coordinates 
in a computational domain of $r_{\rm in} \leq r \leq r_{\rm out}$ 
and $\epsilon \leq \theta \leq \pi -\epsilon$, where $\epsilon$ is set to 
$0.01$ to avoid numerical singularity at poles.
In the radial direction, we set up logarithmically-spaced grids, the number of which is $N_r=512$.
In the polar direction, we use static refinement in order to resolve the scale height of a cold disk,
setting two types of uniformly spaced grids; $\Delta \theta =3(\pi-2\theta_0)/N_\theta$ at 
$\theta_0 \leq \theta \leq \pi-\theta_0$ (near the mid-plane) and 
$\Delta \theta =3(\theta_0-\epsilon)/N_\theta$ otherwise (near the poles).
We set $N_\theta =384~(=128\times 3)$ and $\theta_0=1.3$ so that the disk scale height is sufficiently resolved.
As our fiducial case, we set $r_{\rm in}=1.7\times 10^{-2}~R_{\rm B}$ and $r_{\rm max}=33~R_{\rm B}$.
The grid structure of our simulations is illustrated in Fig. \ref{fig:ini}.

As initial conditions, we adopt a rotational equilibrium distribution ($v_r=v_\theta=0$) 
with a constant specific angular momentum of $j_\infty$:
\begin{equation}
\rho=\rho_\infty\left[1+(\gamma-1)\frac{GM_\bullet}{c_\infty^2 r} 
- \frac{(\gamma-1)}{2}\frac{j_\infty^2}{c_{\rm \infty}^2\varpi^2} \right]
^{1/(\gamma-1)}
\label{eq:torus}
\end{equation}
\citep{Fishbone_Moncrief_1976,Papaloizou_Pringle_1984},
where $\varpi = r\sin \theta $ is the cylindrical radius, $\rho_\infty$ is the ambient gas density,
and $c_\infty$ is the sound speed of the gas.
The first and second terms on the right-hand side present density enhancement via gravity of the central BH
inside the Bondi radius.
The third term expresses the centrifugal force for a given $j_\infty$,
which leads to a maximum value of the density at $r=R_{\rm C}[=j^2_\infty/(GM_\bullet)]$ and $\theta =\pi/2$,
\begin{equation}
\rho_{\rm cr}=\rho_\infty \left(1+\frac{\gamma-1}{2\beta}\right)^{1/(\gamma-1)},
\end{equation}
where $\beta$ is defined by the ratio of the centrifugal radius and Bondi radius,
\begin{equation}
\beta= \frac{R_{\rm C}}{R_{\rm B}}=\frac{j_\infty^2c_\infty^2}{G^2 M_\bullet ^2}.
\label{eq:beta}
\end{equation}
Note that the corresponding specific angular momentum is given by $j_\infty=\sqrt{\beta}R_{\rm B}c_\infty$.
Without viscosity, the density never exceeds this value because of the centrifugal barrier.
In other words, a high-density region with $\rho>\rho_{\rm cr}$ must be formed by 
angular momentum transport due to viscosity.
In Fig. \ref{fig:ini}, the density distribution of the equilibrium torus for $\beta=0.1$ is shown.

Although we employ static mesh refinement near the equatorial region, a disk scale height cannot be resolved well 
if the gas temperature drops below $\sim 0.01~T_\infty$, where $T_\infty$ is the temperature of the ambient gas.
In order to avoid this issue, we set a minimum temperature below which radiative cooling turns off, i.e., 
$\Lambda(T\leq T_{\rm min})=0$.
We adopt $T_{\rm min}=0.1~T_\infty$ as our fiducial value and study two cases with $T_{\rm min}=0.2~T_\infty$ and $0.05~T_\infty$.
As discussed in \S\ref{sec:QQQ}, the Toomre parameter depends on the choice of $T_{\rm min}$ somewhat sensitively.
We will conduct an extrapolation of our results for lower values of $T_{\rm min}\sim 10^4~\K ~(\simeq 10^{-3}~T_\infty)$
and construct the global structure of the accretion disk.

We adopt the outflow boundary condition at the innermost/outermost grid zones from \cite{Stone_Norman_1992},
where zero gradients cross the boundary are imposed on physical quantities
in order to avoid spurious reflection of wave energy at the boundary.
In addition, we impose $v_r\leq 0$ at the inner boundary 
(i.e., inflowing gas from ghost cells is prohibited).
At the poles, the reflective condition is imposed on
the circumferential component of the velocity $v_\theta$.
After the transition to a cold disk, we assume an isothermal Keplerian disk in the inner ghost grids
at $\theta =\theta_{\rm cold}$, where $T(r_{\rm in}, \theta_{\rm cold})\leq 1.2~T_{\rm min}$\footnote{
When the gas density is not high enough to permit cooling, the temperature is higher than the initial value of $T_\infty\gg T_{\rm min}$.};
in practice, the rotational velocity is set to $v_\phi (r<r_{\rm in},\theta_{\rm cold})= v_{\rm Kep}~[1-(5/2)(H/r)^2]^{1/2}$.

\subsection{Basic quantities and parameter choices}

Before we discuss the simulation results, we introduce a dimensionless physical quantity 
which characterizes BH accretion systems.
As a reference of the accretion rate, we define the Bondi accretion rate for adiabatic and fully ionized gas
(i.e., $\gamma=5/3$ and $\mu = 0.62$)\footnote{Note that both of $\gamma$ and $\mu$ are solved self-consistently in our simulations,
but fixed to evaluate the Bondi rate.},
\begin{align}
\dot{M}_{\rm B}&\equiv 4\pi q(\gamma) \rho_\infty 
\frac{G^2M_\bullet ^2}{c_{\infty}^3},\nonumber\\
&\simeq 8.5\times 10^{-3}~\rho_{-22} M_8^2 T_7^{-3/2}~\msunyr,
\end{align}
where $q(\gamma)=1/4$ for $\gamma =5/3$, $\rho_{-22}=\rho_\infty/(10^{-22}~{\rm g~\cc})$,
 $M_8=M_\bullet/(10^8~\msun)$ and $T_7=T_\infty/(10^7~\K)$.
The accretion rate normalized by the Eddington rate is given by
\begin{equation}
\dot{m}_{\rm B}\equiv \frac{\dot{M}_{\rm B}}{\dot{M}_{\rm Edd}}
\simeq 3.7\times 10^{-3}\rho_{-22}M_{8}T_{7}^{-3/2},
\end{equation}
where $\dot{M}_{\rm Edd}(\equiv 10~L_{\rm Edd}/c^2) = 2.3~M_8~\msunyr$ 
is the Eddington accretion rate with a $10\%$ radiative efficiency.

In this work, we set the BH mass to $M_\bullet = 10^8~\msun$, the temperature of 
the ambient gas to $T_\infty=10^7~\K$ ($c_\infty=470~\kms$) at the initial state and 
$\beta =0.1$ and $\alpha_0=0.01$ as our fiducial case.
We study the effects of radiative cooling and heating, varying the gas density $\rho_\infty$ and thus $\dot{m}_{\rm B}$.

\section{Results}
\label{sec:result}

\begin{figure}
\begin{center}
\includegraphics[width=83mm]{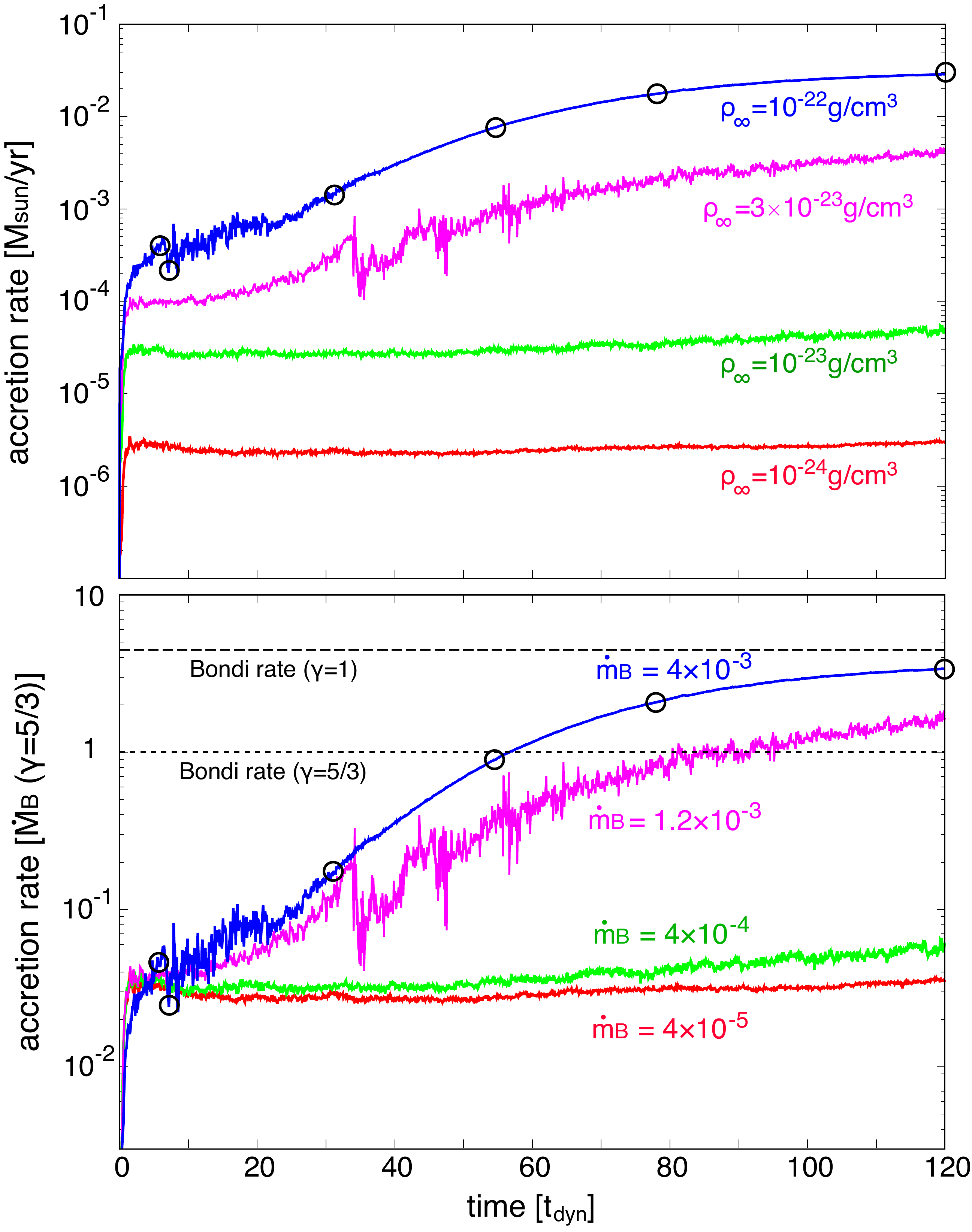}
\caption{Time evolution of the accretion rates
for four different values of $\rho_\infty= 10^{-24}$ (red), $10^{-23}$ (green), $3\times 10^{-23}$ (magenta) and $10^{-22}~\gcc$ (blue), 
which correspond to $\dot{m}_{\rm B} =4\times 10^{-5}$, $4\times 10^{-4}$, $1.2\times 10^{-3}$ and $4\times 10^{-3}$.
The horizontal lines show the Bondi accretion rate for adiabatic gas with $\gamma=5/3$ (short dashed) and for isothermal gas with 
$\gamma=1$ (long dashed).
The accretion rates are shown in units of $\msunyr$ (top) and are normalized by the Bondi rates with $\gamma=5/3$ (bottom).
While for lower densities ($\dot{m}_{\rm B}\la 10^{-3}$), the accretion rates are significantly reduced from the Bondi rate,
for higher densities ($\dot{m}_{\rm B}\ga 10^{-3}$), 
the accretion rate continues to increase and ultimately reaches the Bondi rate for isothermal gas, which is $e^{3/2}(\simeq 4.48)$ 
times higher than that for adiabatic gas (note that for the case with $\dot{m}_{\rm B}=1.2\times 10^{-3}$, 
a steady state is realized at $t\ga 230~t_{\rm dyn}$.).
Open circles mark the epochs at which the radial and angular structure of the density and temperature are shown in 
Figs. \ref{fig:r_rho_temp} and \ref{fig:th_rho_temp}.
}
\label{fig:1}
\end{center}
\end{figure}

Fig. \ref{fig:1} shows the time evolution of gas accretion rates through the inner-most grid 
($r=r_{\rm min}$) for four different values of $\rho_\infty= 10^{-24}$ (red), $10^{-23}$ (green), $3\times 10^{-23}$ (magenta)
and $10^{-22}~\gcc$ (blue), each of which corresponds to $\dot{m}_{\rm B}\simeq 4\times 10^{-5}$, 
$4\times 10^{-4}$, $1.2\times 10^{-3}$ and $4\times 10^{-3}$.
The accretion rates are shown in units of $\msunyr$ (top) and are normalized by 
the Bondi accretion rate with $\gamma=5/3$ for each case (bottom).
The simulation time is normalized by the dynamical timescale at the Bondi radius,
$t_{\rm dyn}(\equiv R_{\rm B}/c_\infty)\simeq 4.1\times 10^3~{\rm yr}~ M_8 T_7^{-3/2}$.
For the lowest density ($\rho_{-22}=0.01$ and $\dot{m}_{\rm B}\simeq 4\times 10^{-5}$), 
the accretion rate increases and saturates at $t\simeq 2~t_{\rm dyn}$.
The saturated value is as small as $\simeq 0.02~\dot{M}_{\rm B}$, which means that the BH accretion rate is 
significantly reduced from the Bondi rate.
As long as $\dot{m}_{\rm B}\la 10^{-3}$, where the accreting gas is adiabatic,
the accretion rate is simply proportional to the ambient gas density.
Therefore, the normalized accretion rate behaves in a self-similar way.

On the other hand, for the highest density ($\rho_{-22}=1$ and $\dot{m}_{\rm B}=4\times 10^{-3}$), 
the self-similar behavior of the accretion rates is no longer valid.
The gas accretion rate further increases even after a few dynamical timescales and ultimately reaches the Bondi rate 
for isothermal gas ($\gamma=1$), which is $e^{3/2}(\simeq 4.5)$ times higher than the Bondi rate for $\gamma=5/3$
(note that for the case with $\dot{m}_{\rm B}=1.2\times 10^{-3}$, a steady state is realized at $t\ga 230~t_{\rm dyn}$).
This result clearly shows that radiative cooling boosts the normalized accretion rate by three orders of magnitude.
As a result, the radiative luminosity from the nuclear regions is also enhanced by many orders of magnitude,
$L/L_{\rm Edd}\simeq 6.3\times 10^{-3}$ (see discussion in \S\ref{sec:mdot-lumbol}).

\begin{figure}
\begin{center}
\includegraphics[width=83mm]{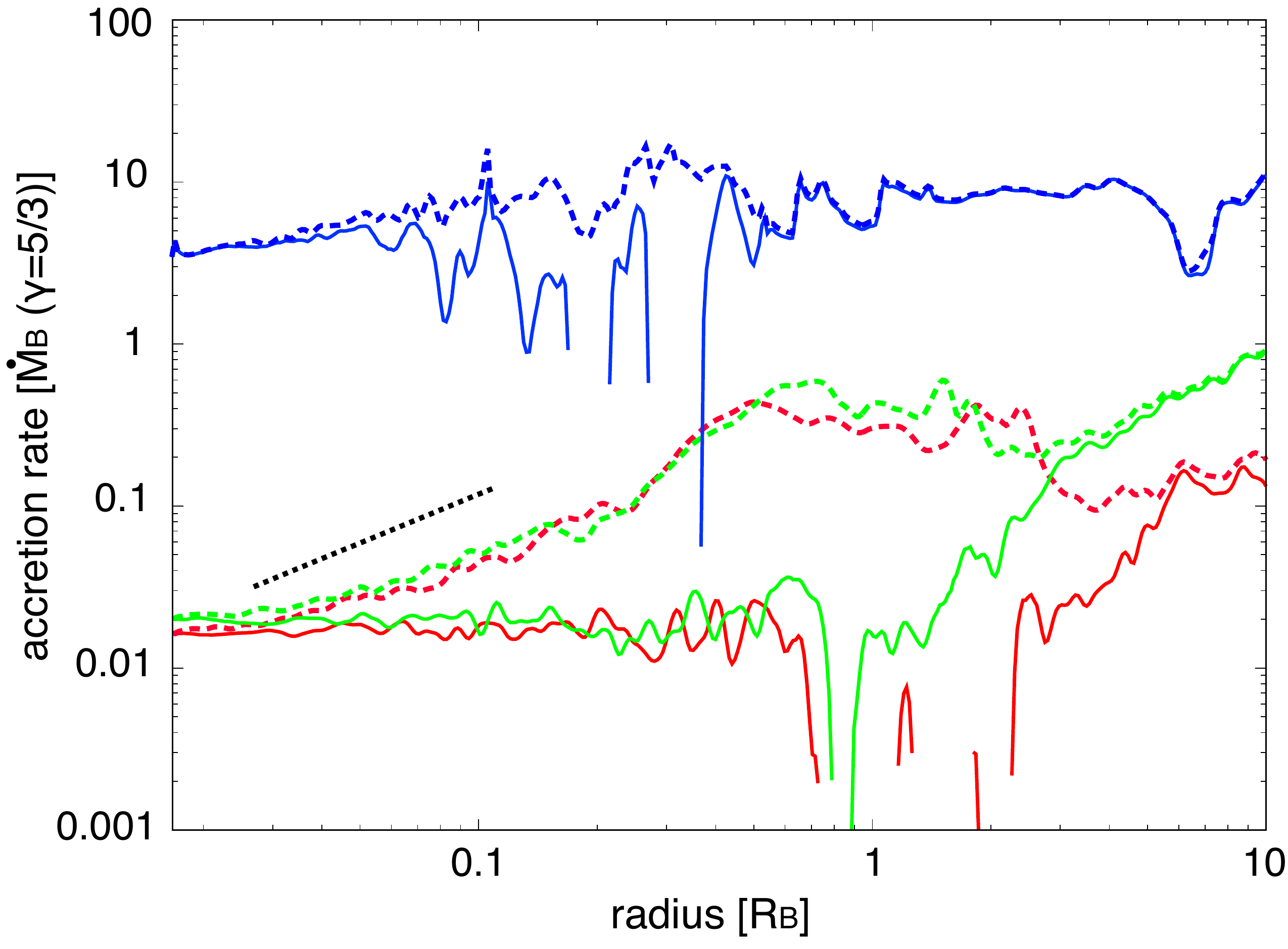}
\caption{Radial structures of the inflow rates (dashed) and net accretion rates (solid) for 
$ \dot{m}_{\rm B}\simeq 4\times 10^{-5}$ (red), $4\times 10^{-4}$ (green), and
$4\times 10^{-3}$ (blue).
The accretion rates are normalized by the Bondi accretion rates for adiabatic gas with $\gamma=5/3$.
The elapsed times correspond to those when the accretion flows are in steady states ($t = 160~t_{\rm dyn}$).
For lower-density cases, the inflow rates decrease toward the center, following $\dot{M}_{\rm in} \propto r$ (black dotted line).
The net accretion rates become a constant value of $\simeq 0.02~\dot{M}_{\rm B}$ within the Bondi radius.
For the highest-density case, the net accretion rate is almost comparable to the inflow rate and
depends on the radius weakly ($\dot{M}_{\rm in}\propto r^p$; $p \simeq 0.3$ at $r<R_{\rm B}$).
}
\label{fig:2}
\end{center}
\end{figure}


\begin{figure*}
 \begin{minipage}{0.49\hsize}
  \begin{center}
   \includegraphics[width=73mm]{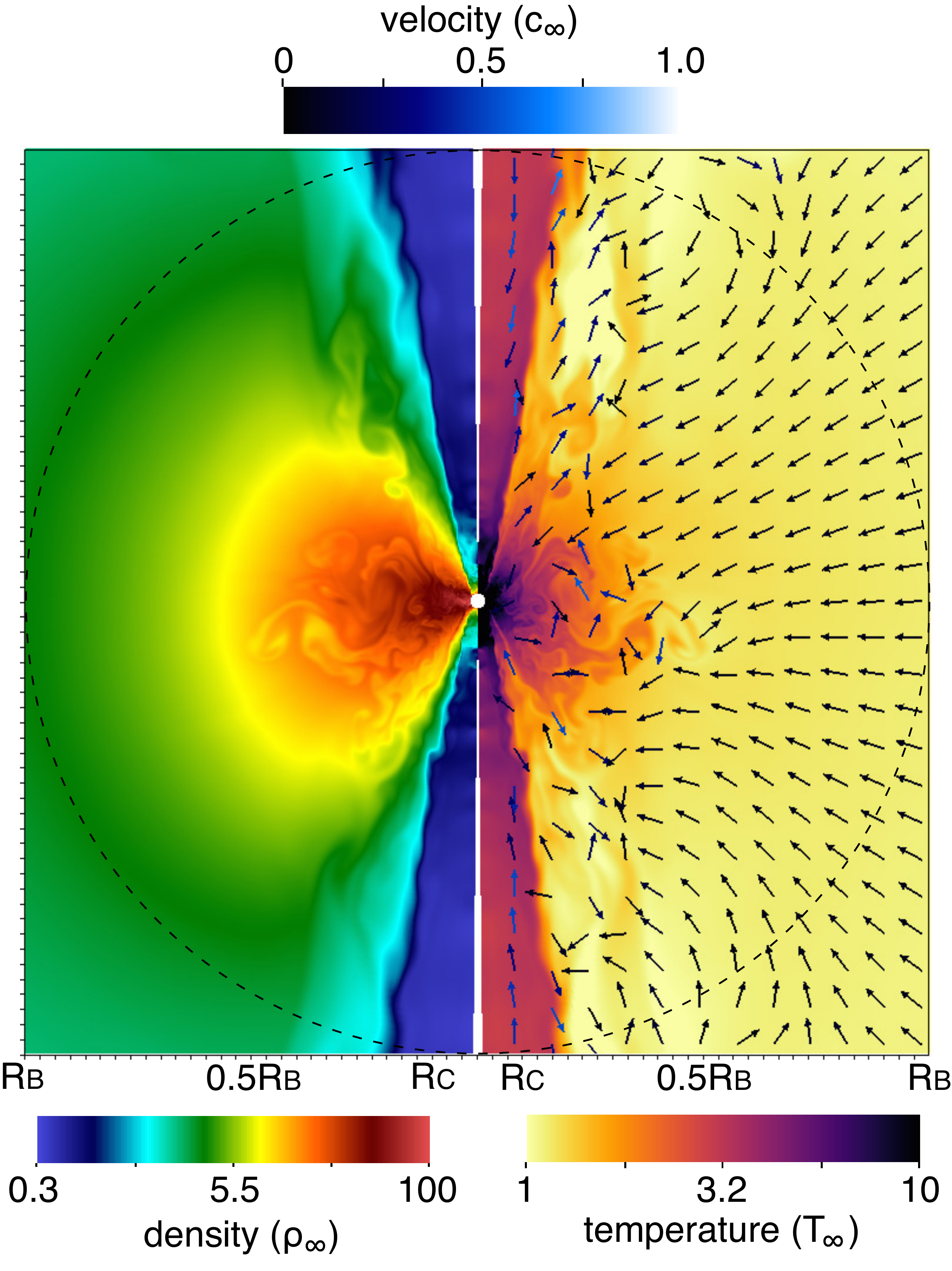}
  \end{center}
 \end{minipage}
 \begin{minipage}{0.49\hsize}
  \begin{center}
   \includegraphics[width=73mm]{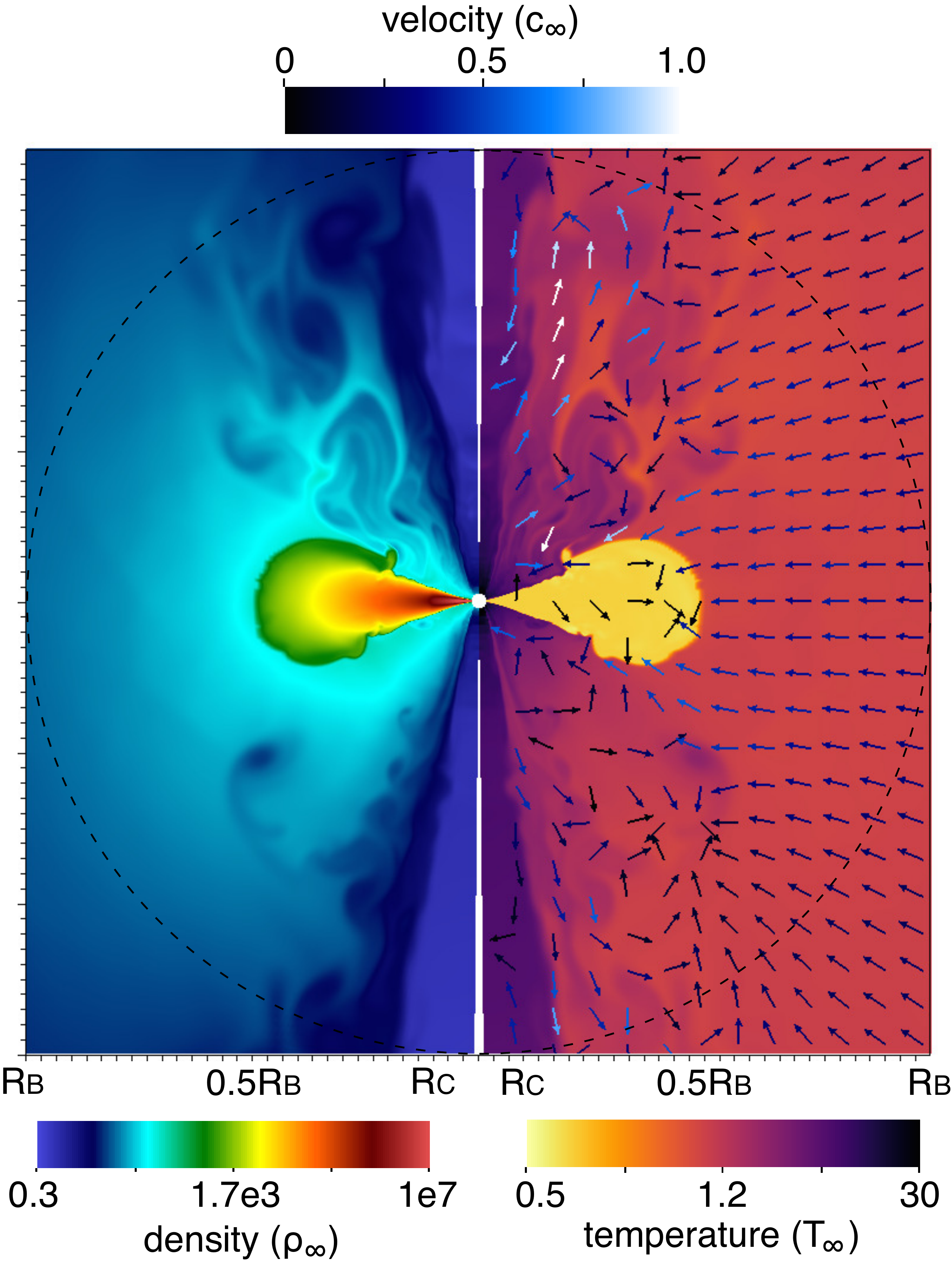}
  \end{center}
 \end{minipage}
 \vspace{1mm}
 \caption{
 Distribution of the gas density (left) and temperature (right) for the case with the highest density,
 $\dot{m}_{\rm B}\simeq 4\times 10^{-3}$.
 At the early stage (left panel), the accretion flow is highly turbulent and forms a geometrically-thick disk 
 inside the Bondi radius.
 In the late stage (right panel), the accreting gas begins to cool, collapses towards the equator and 
 forms a geometrically-thin disk. 
 }
 \label{fig:cont}
\end{figure*}

\begin{figure}
\begin{center}
\includegraphics[width=83mm]{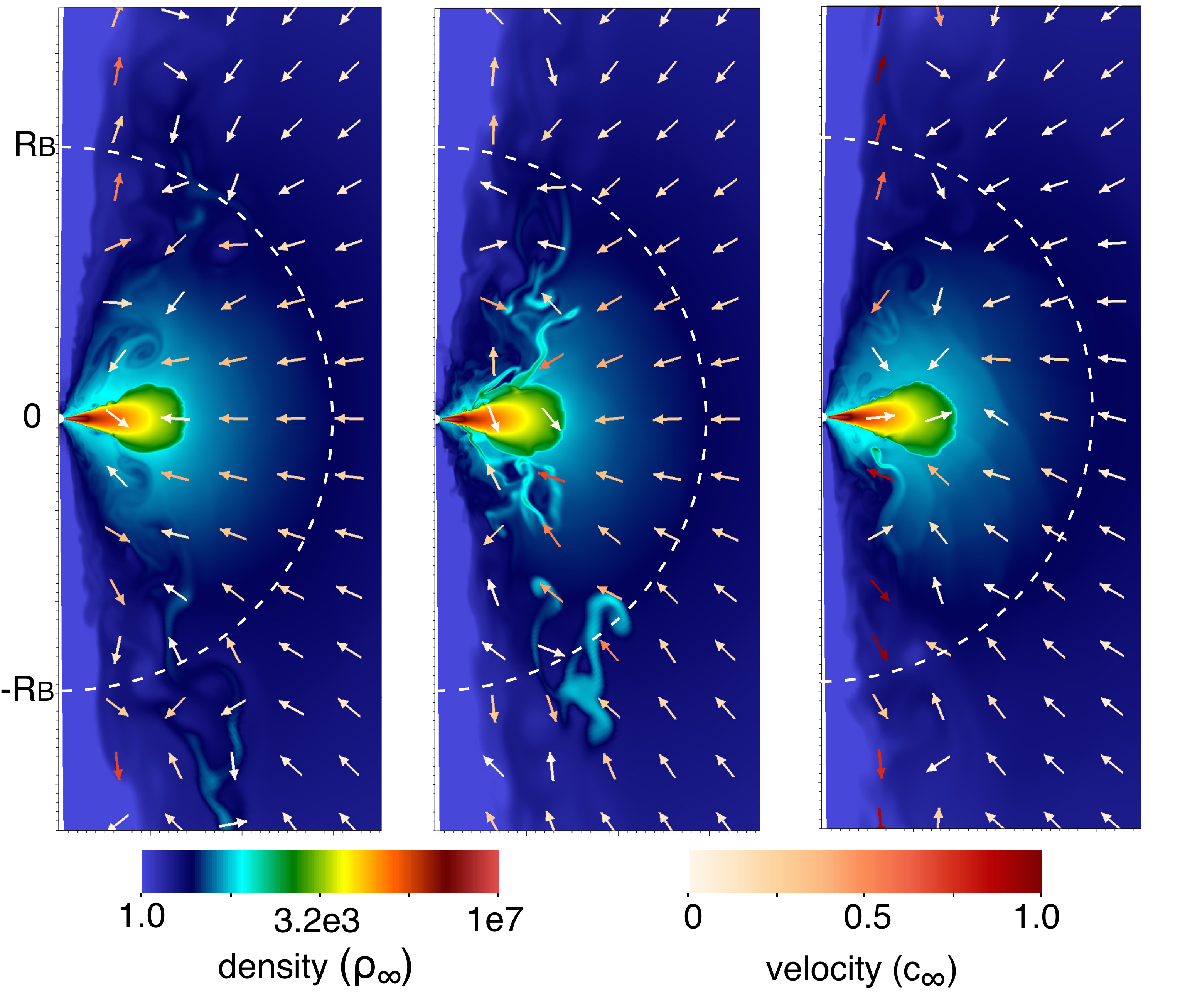}
\end{center}
\vspace{-2mm}
\caption{Density distribution and velocity vectors at three elapsed times around $t\simeq 120~t_{\rm dyn}$
over a short duration with $\Delta t \la 4~t_{\rm dyn}$ for the model with $\dot{m}_{\rm B} \simeq 4\times 10^{-3}$.
A significant fraction of outflowing material to the polar regions cannot escape and
forms clumps around $\sim R_{\rm B}$ due to compression with the ambient gas (left).
Those clumps are captured by the gravity of the BH and accrete onto the disk (middle).
Outflows in the bipolar directions are produced by energy release from the disk and BH.}
 \label{fig:clump}
\end{figure}

\begin{figure}
\begin{center}
\includegraphics[width=78mm]{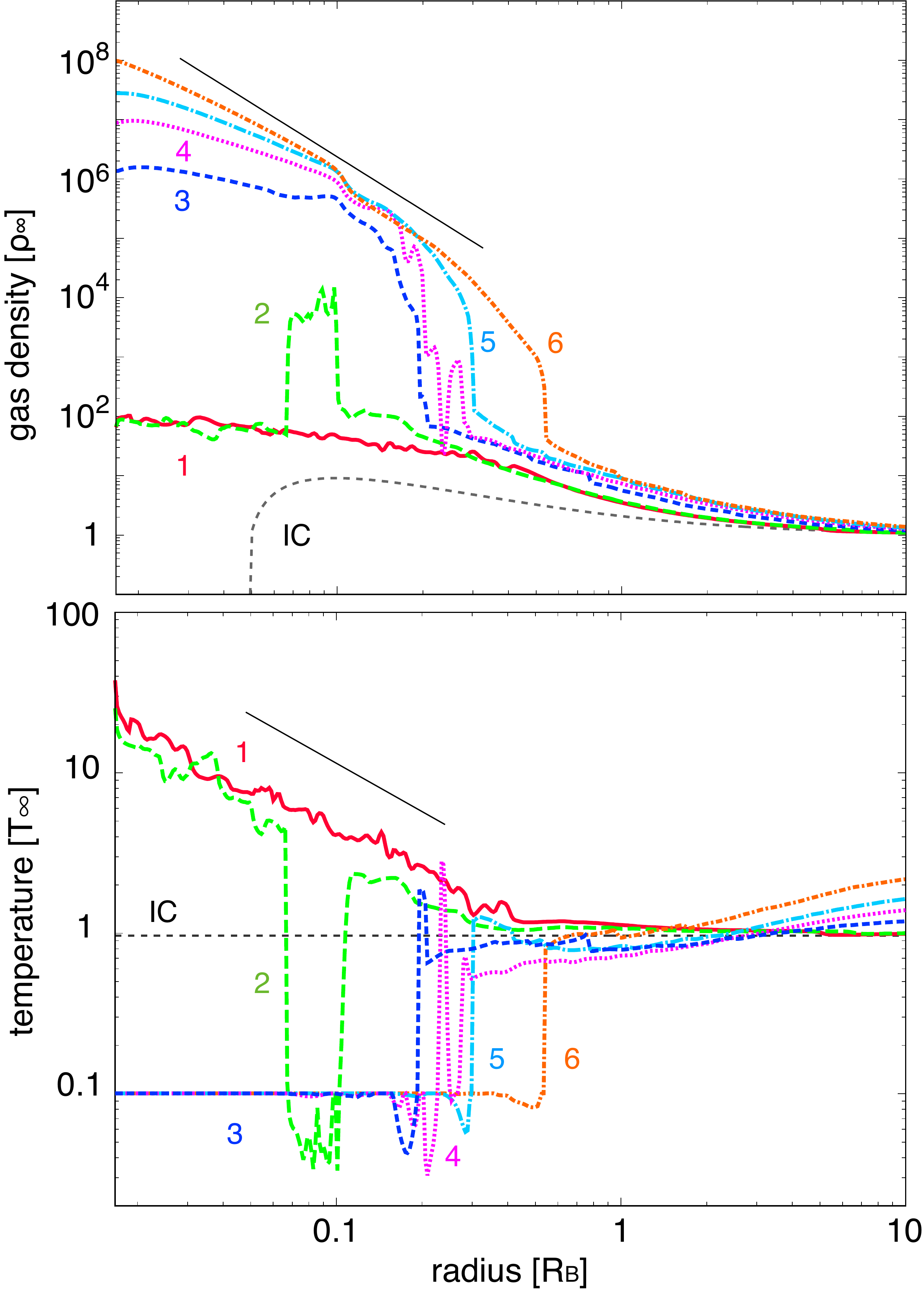}
\caption{
Radial profile of the gas density (top) and temperature (bottom) for the case with the highest density,
$\dot{m}_{\rm B}\simeq 4\times 10^{-3}$.
The curves show the profiles at $t/t_{\rm dyn}=0$ (black dashed), $5.5$ (red), $7.0$ (green), 
$31$ (blue), $55$ (magenta), $78$ (cyan), and $120$ (orange).
}
\label{fig:r_rho_temp}
\end{center}
\end{figure}

\begin{figure}
\begin{center}
\includegraphics[width=78mm]{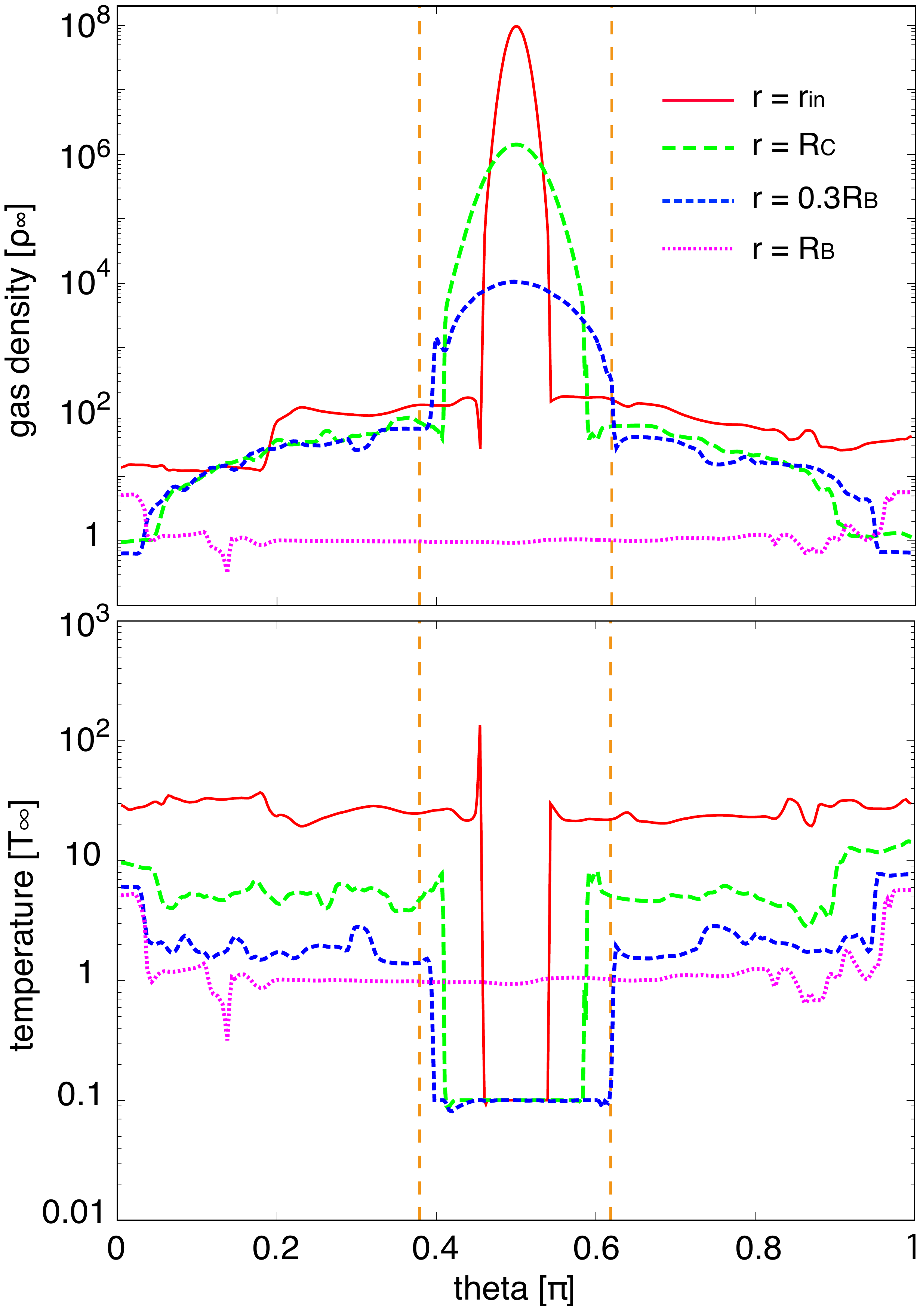}
\caption{
Angular profile of the gas density (top) and temperature (bottom) for the case with the highest density,
$\dot{m}_{\rm B}\simeq 4\times 10^{-3}$ in the final time step ($t=120~t_{\rm dyn}$).
Different curves show the profiles at different radial positions of $r_{\rm in}\leq r \leq R_{\rm B}$.
The cold disk is located within two vertical lines (orange) where grids in the polar direction are refined.
}
\label{fig:th_rho_temp}
\end{center}
\end{figure}

Fig. \ref{fig:2} presents the radial structure of the angle-integrated mass inflow (dashed) and 
net accretion (solid) rates for three different values of $\rho_\infty$ ($4\times 10^{-5} \leq \dot{m}_{\rm B}\leq 4\times 10^{-3}$).
The inflow and outflow rate are calculated as 
\begin{align}
\dot{M}_{\rm in} &= 2\pi r^2 \int_0^\pi \rho ~{\rm min}(v_{\rm r},0)\sin \theta d\theta,\\
\dot{M}_{\rm out} &= 2\pi r^2 \int_0^\pi \rho ~{\rm max}(v_{\rm r},0)\sin \theta d\theta,
\end{align}
and the net accretion rate is defined as $\dot{M}_{\rm net} \equiv -\dot{M}_{\rm in}-\dot{M}_{\rm out}$.
Note that $\dot{M}_{\rm in}<0$ and $\dot{M}_{\rm out}>0$.
For the two cases with lower densities ($\dot{m}_{\rm B} \simeq 4\times 10^{-5}$ and $4\times 10^{-4}$), 
the inflow rates decrease towards the center, following $\dot{M}_{\rm in}\propto r^p$ 
(red and green dashed curves).
The power-law index is consistent with that in CDAFs ($p \simeq 1$).
This fact indicates that convective motion driven by viscous heating suppresses the gas inflow 
(e.g., \citealt{Narayan_2000, Quataert_2000}, see also \citealt{Inayoshi_2018} for details).
The net accretion rates are almost constant at $\sim 0.02~\dot{M}_{\rm B}$ within the Bondi radius.
For the highest-density case ($\dot{m}_{\rm B}\simeq 4\times 10^{-3}$), the inflow rate becomes comparable 
to the net accretion rate within the centrifugal radius.
As shown in Fig \ref{fig:1}, the net accretion rate, normalized by the Bondi rate for $\gamma =5/3$, is two orders of magnitude 
higher than those in the lower-density cases.
The radial dependence of the inflow rate becomes flatter ($p \sim 0.3$) because convective motion ceases due to efficient radiative cooling.

In Fig. \ref{fig:cont}, we present the distribution of the gas density (left) and temperature (right) at two different elapsed times
for the run with $\dot{m}_{\rm B}=4\times 10^{-3}$.
The velocity vectors are over plotted in the temperature distribution.
At the early stage of $t\la 10~t_{\rm dyn}$ (left panel), the accretion flow is highly turbulent (but subsonic) 
and forms a geometrically-thick torus inside the Bondi radius.
Since the accreting gas is still adiabatic, the structure of the accretion flow is similar to that of a CDAF.
We note that 3D simulations of non-axisymmetric accretion flows performed by \cite{Igumenshchev_2000} 
suggest that the results are essentially similar to those obtained in 2D simulations, and the turbulent eddies 
are nearly axisymmetric and transport angular momentum inwards. 
This allows us to justify that our 2D simulations can capture the essential physics.
At the late stage of $t\ga 10~t_{\rm dyn}$ (right panel), the accreting gas begins to cool, collapse towards the equator and forms a geometrically-thin disk.
The disk increases its mass via gas supply from the Bondi radius, and feeds the central BH with the aid of viscous processes
transporting the angular momentum outwards.
Since radiative cooling is inefficient above the disk, the cold disk is embedded in hot gas ($T \gg T_\infty$).
As shown in Fig \ref{fig:1}, the accretion flow settles into a steady state at $t\ga 100~t_{\rm dyn}$,
where the BH accretion rate is as high as $\simeq 0.02~\dot{M}_{\rm Edd}$.
In the steady state, the radiative luminosity from the BH ($r<r_{\rm in}$) is on the order of $\sim 0.01~L_{\rm Edd}$.
The combination of heating and outward momentum caused by radiation produces weak outflows at $r\la R_{\rm B}$.
A small fraction of the gas can escape from the Bondi radius with a velocity of $v_{r}\simeq 2~c_\infty$, 
but the mass outflow does not reduce the BH accretion rate significantly.

In Fig. \ref{fig:clump}, we show the density distribution and velocity vectors at three elapsed times around $t\simeq 120~t_{\rm dyn}$
over a short duration with $\Delta t \la 4~t_{\rm dyn}$ for the model with $\dot{m}_{\rm B} \simeq 4\times 10^{-3}$.
Since the outflow velocity is comparable to the sound speed at the Bondi radius, a large fraction of the outflowing gas to
the polar regions cannot escape outside the Bondi radius.
Instead, the gas is compressed by the ambient gas and forms clumps around $\sim R_{\rm B}$.
The clumps are captured by the gravity of the BH and accrete back onto the disk again. 
Such a burst of gas accretion increases the energy output from the disk (viscous dissipation) and BH (radiation).
As a result, mass outflows are launched to the bipolar directions episodically.

Fig. \ref{fig:r_rho_temp} shows the radial profiles of the gas density (top) and temperature (bottom)
at the mid-plane for $\dot{m}_{\rm B}\simeq 4\times 10^{-3}$ at different elapsed times, 
which correspond to filled circles shown in Fig. \ref{fig:1}.
At the early stage (red curves, epoch 1), the gas density and temperature follow $\rho \propto r^{-1/2}$ and $T\propto r^{-1}$
within the centrifugal radius, respectively.
The radial structures are consistent with those of a CDAF solution \citep[e.g.,][]{Quataert_2000}.
At the late stages, radiative cooling induces gas collapse onto the mid-plane.
This transition begins to occur around the centrifugal radius (see green curves, epoch 2), 
where the cooling condition is satisfied first in the flow ($t_{\rm cool}<t_{\rm th, vis}$; see \S\ref{sec:vis_cool}).
After the transition, the gas density in the disk dramatically increases to $\simeq 10^8~\rho_\infty$ at $r\simeq r_{\rm in}$, 
and the disk outer edge moves outwards by angular momentum transport (see epoch $3-6$).
The gas temperature drops to the threshold value we set ($T_{\rm min}=0.1~T_\infty$) via optically-thin cooling.
Note that this transition occurs in an unstable way because the cooling timescale becomes shorter with decreasing temperature
as $t_{\rm cool}\propto T^s$, where $s \simeq 0.5-1.5$ at $T\la 10^7~\K$ due to free-free emission, metal lines and recombination.
In the steady state, the density profile at the mid-plane follows
\begin{align}
\rho &= \frac{\dot{M}_{\rm d}}{6\pi \alpha}\frac{GM_\bullet}{r^3 c_s^3},\nonumber\\
&=\frac{e^{3/2}}{6\alpha}~\rho_\infty
\left(\frac{c_{\infty}}{c_{\rm s}}\right)^3
\left(\frac{r}{R_{\rm B}}\right)^{-3},
\label{eq:rho_d}
\end{align}
where $\dot{M}_{\rm d}$ is the accretion rate through the disk, comparable to the Bondi rate for isothermal gas,
and $c_{\rm s}$ is the sound speed of the gas in the disk with a temperature of  $T\simeq T_{\rm min}$.
The solid line in the top panel shows Eq. (\ref{eq:rho_d}) at $r\leq R_{\rm C}(=0.1~R_{\rm B})$.
The temperature exterior to the cold disk, on the other hand, becomes hotter due to Compton heating
($T_{\rm C}=10^8~\K$) as the BH feeding rate increases.

In Fig. \ref{fig:th_rho_temp}, we present the angular profiles of the density (top) and temperature (bottom) 
as a function of the polar angle $\theta$, at radial positions of $r_{\rm in}\leq r \leq R_{\rm B}$.
Inside the Bondi radius, the density increases toward the center and the mid-plane ($\theta=\pi/2$).
Within the cold disk ($0.4\la \theta/\pi \la0.6$), the temperature is constant at $T=T_{\rm min}(=0.1~T_\infty)$.
Above the disk, the gas is heated up with decreasing radii.
Note that the cold disk region is located within the region where grids in the polar direction are refined ($\theta_0 \leq \theta \leq \pi-\theta_0$).

\section{Physical interpretation}
\label{sec:ana}

\subsection{The formation of a cold accretion disk}\label{sec:vis_cool}

\begin{figure}
\begin{center}
\includegraphics[width=83mm]{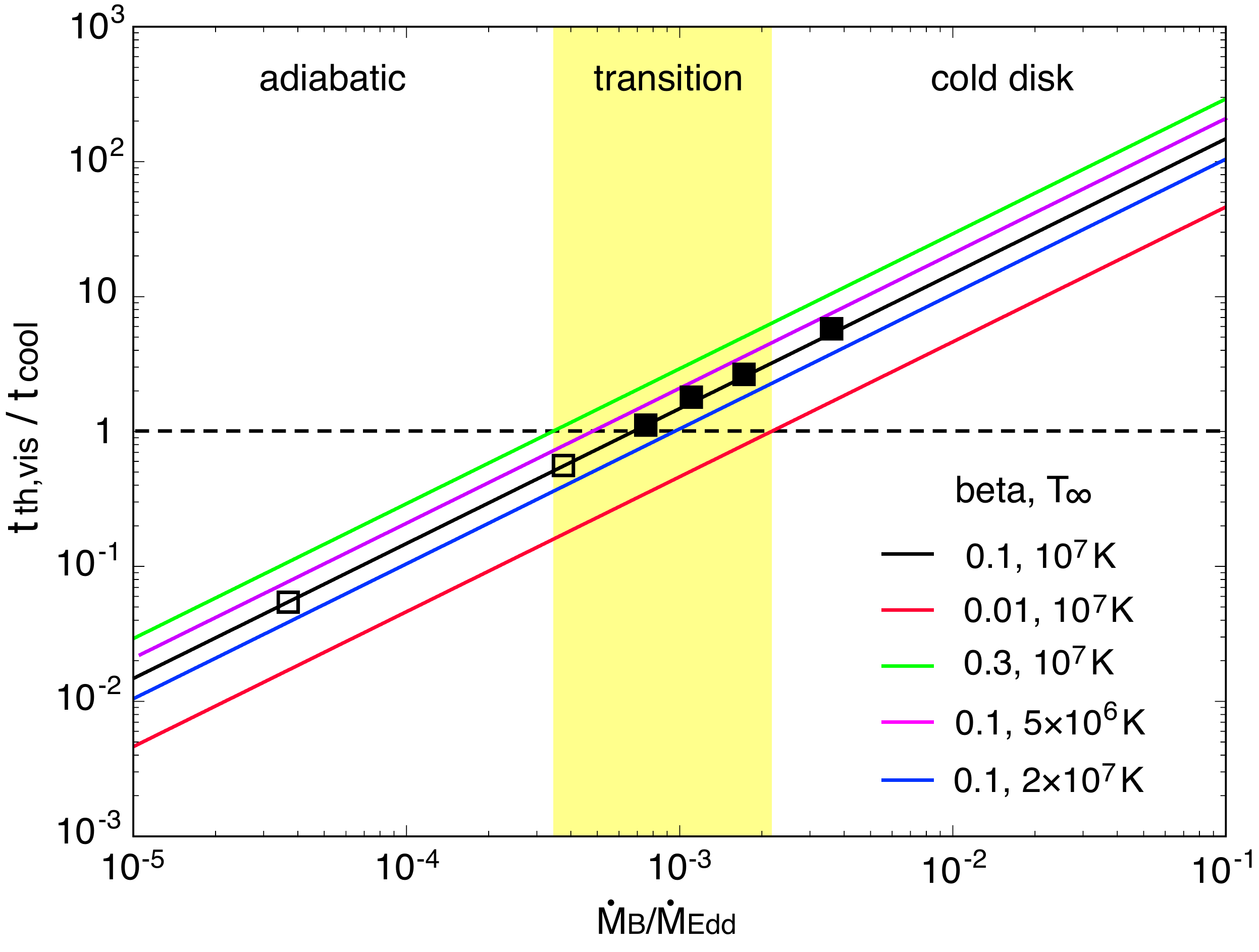}
\vspace{-1mm}
\caption{The ratio of the viscous heating timescale $t_{\rm th,vis}$ to the radiative cooling timescale $t_{\rm cool}$ 
as a function of $\dot{M}_{\rm B}/\dot{M}_{\rm Edd}$, for different values of the angular momentum ($\beta$)
and ambient gas temperature ($T_\infty$).
The shaded region at $4\times 10^{-4}\la \dot{M}_{\rm B}/\dot{M}_{\rm Edd}\la 2\times 10^{-3}$ presents the transition region 
from adiabatic flows to cold disk accretion.
For $\beta = 0.1$ and $T_\infty =10^7~\K$, we show our numerical results where adiabatic accretion is maintained (open) and 
where cold accretion disk forms (filled).
}
\label{fig:cool_vis}
\end{center}
\end{figure}

For lower densities at the Bondi radius ($\dot{M}_{\rm B}/\dot{M}_{\rm Edd}\ll 10^{-3}$),
the accreting gas onto a BH forms a geometrically-thick torus.
Namely, the gas density, temperature and inflow rate are consistent with those of a CDAF solution: 
$\rho \propto r^{-1/2}$, $T\propto r^{-1}$ and $\dot{M}\propto r$ \citep[e.g.,][]{Quataert_2000}.
Inside the torus structure, viscous energy dissipation heats the gas and drives convective motion 
on a thermal timescale of $t_{\rm th,vis}\simeq 1/[\gamma(\gamma-1)\alpha \Omega]^{-1}\propto r^{3/2}$.
With the CDAF solution, the bremsstrahlung cooling rate per volume is 
$Q^-_{\rm br} \propto \rho^2 T^{1/2} \propto r^{-3/2}$ and the cooling timescale is $t_{\rm cool}\propto \rho T/Q^-_{\rm br} \propto r^0$.
Fig. \ref{fig:cool_vis} shows the ratio of $t_{\rm th,vis}/t_{\rm cool}$ as a function of $\dot{m}_{\rm B}$ for different values of 
$\beta (=R_{\rm C}/R_{\rm B})$ and $T_\infty$, 
adopting the radial density and temperature profile of a CDAF solution inside the Bondi radius \citep{Inayoshi_2018}.
The ratio of the two timescales is estimated as 
\begin{equation}
\frac{t_{\rm th,vis}}{t_{\rm cool}}\simeq 1.48~ T_7^{1/2}
\left(\frac{\alpha}{0.01}\right)^{-1}
\left(\frac{\dot{m}_{\rm B}}{10^{-3}}\right)
\left(\frac{\beta}{0.1}\right)^{1/2},
\label{eq:vis_cool}
\end{equation}
where the ratio is evaluated at the outer edge of the torus ($r \simeq 2R_{\rm C}$).
We note that this scaling relation agrees with the results shown in Fig.~\ref{fig:cool_vis}, as long as the angular momentum 
is sufficiently small ($\beta \la  0.3$).
From Eq. (\ref{eq:vis_cool}), we obtain the critical accretion rate above which bremsstrahlung cooling plays 
an important role in the accretion flow as 
\begin{equation}
\frac{\dot{M}_{\rm B}}{\dot{M}_{\rm Edd}}\ga \dot{m}_{\rm cool}\equiv 6.8\times 10^{-4}~T_7^{-1/2}
\left(\frac{\alpha}{0.01}\right)
\left(\frac{\beta}{0.1}\right)^{-1/2}.
\label{eq:crit_mdot}
\end{equation}
In Fig. \ref{fig:cool_vis}, we also present the results of numerical simulations where a cold accretion disk forms (filled squares) 
and those where the accretion flow is adiabatic (open squares).
For a wide range of the angular momentum ($0.01\leq \beta \leq 0.3$) and ambient gas temperature ($5\times 10^6 \leq T_\infty/\K \leq 10^7$), 
the transition from adiabatic accretion flows to cold disk accretion occurs at $4\times 10^{-4} \la \dot{m}_{\rm cool}\la 2\times 10^{-3}$.
We note that Eq. (\ref{eq:crit_mdot}) is not valid for lower temperatures where metal-line cooling dominates bremsstrahlung cooling significantly
($T_\infty \ll 5\times 10^6~\K$).

It is worth comparing our results to those obtained in previous work which investigated the radiative cooling effects
in hot and optically-thin accretion flows.
Early studies by \cite{Yuan_2001,Yuan_2003} have discussed the properties of luminous hot accretion flows (LHAFs)
in the inner region ($r\la 10^2~R_{\rm Sch}$), where the sum of viscous heating and compressional heating balance 
with cooling via Coulomb collisions between ions and electrons.
For higher rates of $\dot{M}\ga \dot{M}_{\rm LHAF}\simeq 1.5~\alpha^{0.7}\dot{M}_{\rm Edd}$\footnote{
The critical rate would depend on the choice of the outer edge of the disk. 
For $R_{\rm out}=10^4~R_{\rm Sch}$, \cite{Yuan_2001} obtained $\dot{M}_{\rm LHAF}/\dot{M}_{\rm Edd}\simeq 0.25$, 
$5\times 10^{-2}$ and $5\times 10^{-4}$ for $\alpha=0.1$, $10^{-2}$ and $10^{-3}$, respectively.},
hot accretion flows no longer exist because radiative cooling efficiently carries the energy away.
We note that the critical value of $\dot{M}_{\rm LHAF}$ is two orders of magnitude higher than the critical rate given by 
Eq. (\ref{eq:crit_mdot}) assuming $\alpha=0.01$.
Another qualitative difference is that accreting gas from the Bondi radius begins to cool at the disk outer radius,
while in LHAFs the inner region of the disk at $<10^2~R_{\rm Sch}$ collapses via cooling.
\cite{Wu_Xie_2016} have also studied the cooling transition of LHAFs at $r\la 10^2~R_{\rm Sch}$, performing hydrodynamical simulations.
They found a critical accretion rate of $\dot{M}_{\rm LHAF} \sim 3\alpha \dot{M}_{\rm Edd}$, above which cold and dense
clumpy and/or filamentary structures are formed within hot gas.
We note that in their simulations, gas supply from larger scales is not considered unlike our simulations.
The different treatment of the outer boundary conditions would affect the cooling conditions
and significantly lower the critical accretion rate given in Eq. (\ref{eq:crit_mdot}).

In this paper, we do not discuss the regime of higher accretion rate where various feedback processes would be important, e.g., radiation, winds and jets
(e.g., \citealt{Sazonov_2005,Ciotti_Ostriker_2001,Proga_2007,Yuan_2009,Ciotti_2009,Booth_Schaye_2009,Milosavljevic_2009,
Yuan_Li_2011,Novak_2011,Choi_2012,Choi_2014,Costa_2014,Xie_2017,Yuan_2018,Yoon_2018,Bu_Yang_2019}, 
and see \citealt{Fabian_2012} references therein).
We define the critical BH accretion rate $\dot{m}_{\rm fb}$, above which a steady state no longer exists due to feedback 
but gas accretion occurs episodically.
The exact value of $\dot{m}_{\rm fb}$ is still uncertain, depending on which feedback processes dominate. 
Instead of discussing each process, we here adopt a critical radiation luminosity of $L_{\rm bol}/L_{\rm Edd}\simeq 0.02$ as a reference
\citep[e.g.,][]{McClintock_Remillard_2006}.
Since the BH accretion rate through a cold (isothermal) disk is $\sim e^{3/2}(\simeq 4.5)$ times higher than the Bondi accretion rate 
estimated with $\gamma=5/3$, the critical value of the Bondi rate is $\dot{m}_{\rm fb,B}\equiv \dot{m}_{\rm fb}/e^{3/2}\simeq 6.7\times 10^{-3}$,
where $L_{\rm bol}(\dot{m}_{\rm fb})=0.02~L_{\rm Edd}$ (see Fig. \ref{fig:summary_result} and discussion in \S\ref{sec:mdot-lumbol}).
We leave the study of the strong feedback domain for future work.

\begin{figure}
\begin{center}
\includegraphics[width=81mm]{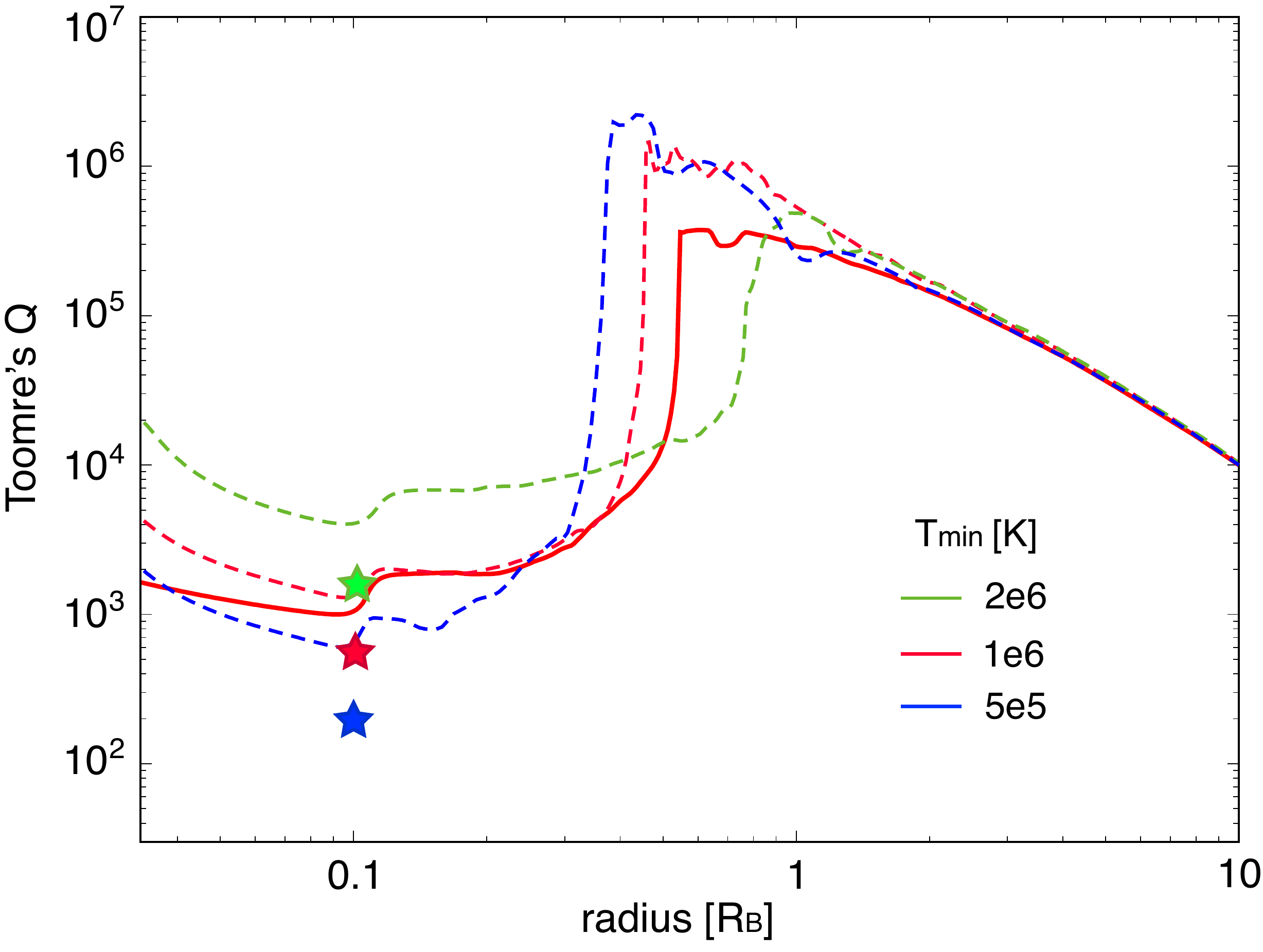}
\caption{
Radial profiles of the Toomre parameter for $\dot{m}_{\rm B}\simeq 4\times 10^{-3}$ at the final time step ($t=120~t_{\rm dyn}$)
with an inner-most radius of $r_{\rm in}/R_{\rm B} = 1.7 \times 10^{-2}$ (fiducial case, solid) and $3.4\times 10^{-2}$ (dashed). 
Each curve presents the profile for a minimum temperature of $T_{\rm min}=2\times 10^6~\K$ (green), $10^6~\K$ (red) 
and $5\times 10^5~\K$ (blue), respectively.
The minimum values are located at $r=R_{\rm C}$. 
Star symbols indicate those shown in Eq. (\ref{eq:QQ}).
}
\label{fig:r_Q}
\end{center}
\end{figure}

\subsection{Toomre instability and an extension of the disk solution towards the BH}\label{sec:QQQ}

We now focus on the intermediate regime of $\dot{m}_{\rm cool} \la \dot{m}_{\rm B} \la \dot{m}_{\rm fb,B}$, 
where a dense and geometrically-thin disk forms due to radiative cooling, but feedback does not affect the BH accretion.
The cold disk may then fragment into clumps and form stars by a spiral-mode gravitational instability, which is characterized by the Toomre parameter
\citep{Toomre_1964}:
\begin{equation}
Q\equiv \frac{c_{\rm s}\Omega}{\pi G \Sigma}.
\end{equation}
Assuming a steady state, the gas surface density is estimated as $\Sigma \simeq \dot{M}_{\rm d}/(3\pi \nu)$,
where $\dot{M}_{\rm d}$ is the gas accretion rate through the disk.
Thus, we obtain 
\begin{align}
Q&=\frac{3\alpha c_{\rm s}^3}{G \dot{M}_{\rm d}},\label{eq:QQ}
\\ 
&\simeq 2.1~T_{\rm min,4}^{3/2}M_8^{-1}
\left(\frac{\alpha}{0.01}\right)
\left(\frac{\dot{m}_{\rm B}}{10^{-3}}\right)^{-1},\nonumber
\end{align}
where $T_{\rm min,4}=T_{\rm min}/(10^4~\K)$.
Fig. \ref{fig:r_Q} shows the radial profile of the Toomre parameter for $T_{\rm min}=0.1~T_\infty~(=10^6~\K)$ (red solid curve).
In the cold disk region, the Toomre'parameter is almost constant as a function of radius
and has a minimum value of $Q\simeq 10^3$ at the centrifugal radius ($R_{\rm C}=0.1~R_{\rm B}$).
This result agrees with the analytical estimate shown by the star symbol (see Eq. \ref{eq:QQ}) within a factor of two.
Dashed curves present the cases for $T_{\rm min}=10^6~\K$ (red), $2\times 10^6~\K$ (green) 
and $5\times 10^5~\K$ (blue), but with an inner-most radius of $r_{\rm in}=3.4\times 10^{-2}~R_{\rm B}$,
which is twice larger than our fiducial case (solid).
The Toomre values at $r=R_{\rm C}$ decreases for lower $T_{\rm min}$, following $Q\propto T_{\rm min}^{3/2}$.
Therefore, we would expect that once radiative cooling leads to the formation of a cold disk and the temperature decreases below $10^4~\K$, 
the disk would become unstable because of its self-gravity.

When the Toomre parameter decreases to $\sim O(1)$, the accretion disk becomes gravitationally unstable \citep{Toomre_1964} and 
induces non-axisymmetric structure like spiral arms transporting angular momentum outward efficiently 
\citep{Gammie_2001,Goodman_2003,Rice_2005,Vorobyov_Basu_2006,Kratter_2008,Machida_2010,Kuiper_2011,Zhu_2012,Takahashi_2016}.
The strength of gravitational torque is characterized by an effective viscosity of $\alpha_{\rm eff} \sim O(1)$.
To mimic this effect, we parameterize the effective viscosity as a function of $Q$ in the following:
\begin{equation}
\alpha_{\rm eff} = \alpha_0 + \alpha_{\rm max}\exp(-Q^A/B),
\label{eq:alphaeff}
\end{equation}
where $\alpha_0\simeq 0.01$ is the viscous parameter induced by the MRI and $\alpha_{\rm max}\sim 0.1-1$ is the maximum viscous strength.
Hydrodynamical simulations of a self-gravitating disk suggest that the disk tends to fragment when the strength of gravitational torque 
exceeds a critical threshold value of $\alpha_{\rm max}$.
The threshold value depends on mass accretion from an infalling envelope, realistic radiative cooling, 
and radiative trapping of energy inside clumps \citep{Zhu_2012}.
In the context of massive star formation, the critical value for fragmentation is suggested to be $\alpha_{\rm max}\sim 0.1-1$ 
\citep[e.g.,][]{Kratter_2008}\footnote{Numerical simulations for isolated disks without infalling flow from larger scales suggest 
lower critical values of $\alpha_{\rm max}\simeq 0.06$ \citep{Rice_2005}.}.
Based on those results, we adopt $\alpha_{\rm max}=1$ as a reference value.
We also note that the choice of the two free parameters of $(A,B)$ hardly affects the strength of gravitational torque
(see Appendix \ref{sec:app1}) and thus they are set to $(A,B)=(10,1)$.
Using Eqs. (\ref{eq:QQ}) and (\ref{eq:alphaeff}), we obtain the critical accretion rate above which the disk becomes gravitationally unstable 
($Q\sim 1$ and $\alpha_{\rm eff}\ga \alpha_0$) as
\begin{equation}
\frac{\dot{M}_{\rm B}}{\dot{M}_{\rm Edd}} \ga 5.6 \times 10^{-3}~T_{\rm min,4}^{3/2}M_8^{-1},
\label{eq:toomre_mdot}
\end{equation}
or
\begin{equation}
\dot{M}_{\rm B} \ga 1.3\times 10^{-2}~T_{\rm min,4}^{3/2}~\msunyr,
\end{equation}
where $\alpha_{\rm eff}\geq 0.03$ is adopted as the threshold value.
The minimum temperature is as high as $T_{\rm min} \simeq 3000~\K$, which is determined by H$^-$ continuum radiation.
Thus, once radiative cooling leads to the formation of a dense disk in the Bondi radius of a BH with $M_\bullet \ga 10^8~\msun$,
it would turn gravitationally unstable immediately.
For less massive BHs with $M_\bullet \sim 10^7~\msun$, accretion rates as high as $\dot{m}_{\rm fb,B}\sim 10^{-2}$ are 
required to trigger the onset of gravitational instability.

As a specific example, we briefly discuss the disk structure of the nuclear BH in M87 galaxy
($M_\bullet=6.15\times 10^9~\msun$ and $\dot{m}_{\rm B}=1.34\times 10^{-3}$; \citealt{Gebhardt_2011,Russell_2015}), 
which satisfies Eq. (\ref{eq:toomre_mdot}).
Therefore, the nuclear accretion disk within the Bondi radius becomes unstable, leading to
efficient fragmentation and star formation \citep[e.g.,][]{Goodman_2003,Tan_Blackman_2005}\footnote{
The nuclear disk in M87 is no longer stable because $\psi <10^{-2}$, where $Q\la 1$ and $\alpha_{\rm eff}\sim \alpha_{\rm max}$ 
(see Appendix \ref{sec:app1}).}.
Since star formation will consume the disk mass and the newly formed stars also will heat the disk,
the remaining gas can be stabilized, result in $Q\sim 1$.
In the marginally unstable disk, the surface density and disk mass are given by
\begin{align}
\Sigma \simeq \frac{c_{\rm s}\Omega}{\pi GQ}\simeq 2.4\times 10^4~{\rm g~cm^{-2}}~ T_{\rm min,4}^{1/2} M_9^{-1}Q^{-1} \hat{r}_3^{-3/2},
\end{align}
and
\begin{align}
M_{\rm d}(<R_{\rm C}) \simeq 9.5\times 10^{6}~\msun ~T_{\rm min,4}^{1/2} M_9 Q^{-1},
\end{align}
where $\beta(=R_{\rm C}/R_{\rm B})=0.1$ is adopted.
A fraction $\epsilon_\star$ of the disk mass would be consumed by star formation.
Assuming the star formation efficiency $\epsilon_\star=0.01$ inferred from the Kennicut-Schmidt law 
\citep{Schmidt_1959,Kennicutt_1998,Krumholz_2005}, 
the total stellar mass is $M_{\star}\sim 10^{6}~\msun$ and the expected number of supernovae (SNe) is 
$N_{\rm SN}\simeq M_{\star}/m_\star\sim 10^4$, where $m_\star \sim 100~\msun$ is the stellar mass required to generate an SN 
for a Kroupa initial mass function \citep{Kroupa_2001}.
Since the total SN energy $N_{SN}E_{\rm SN}\simeq 10^{55}E_{51}$ erg is injected over the lifetime of massive stars
$t_{\rm life}\sim 20$ Myr, thus the kinetic luminosity is $L_{\rm kin}\simeq 2\times 10^{40}~{\rm erg~s}^{-1}$.
Let us assume that super-bubbles expand as pressure-driven snowplows with no radiative cooling in their interior.
Then, the criterion for ``breakout'' from the cold disk is expressed as $L_{\rm kin}\gg \rho_{\rm a} v_{\rm t}^3 A$, 
where $\rho_{\rm a}$ is the ambient gas density, $v_{\rm t}$ is the turbulent velocity in the disk, and $A~(\sim H^2)$ 
is the total surface area of the bubbles \citep{MacLow_McCray_1988,Koo_McKee_1992,Kim_Ostriker_2017}.
In this case for M87 ($\rho_{\rm a}\sim 10^{-16}~\gcc$, $v_{\rm t}\sim c_{\rm s}\sim 10~\kms$ and 
$H\sim 2\times 10^{17}$ cm at $r\simeq R_{\rm C}$), the ratio of the SN kinetic luminosity to that of turbulent gas 
is estimated as $L_{\rm kin}/(\rho_{\rm a} v_{\rm t}^3 A)\sim 5000\gg 1$.
This implies that SN feedback would likely blow the gas away from the cold disk, making star formation episodes brief.

In a giant elliptical galaxy harboring an SMBH with a mass of $M_\bullet \ga 10^9~\msun$ fed at $\dot{m}_{\rm B} \ga \dot{m}_{\rm cool}$,
star formation activities in the nuclear region within the Bondi radius would be ubiquitous \citep[e.g.,][]{Tan_Blackman_2005,Gan_2018}.
In fact, this hypothesis would explain the following observational features of the M87 system: 
the existence of a cold disk in the nuclear region and a low level of star formation.
The cold disk, which is traced by H$\alpha$ emission, has a size of $\sim 20-40~\pc$ \citep[e.g.,][]{Harms_1994, Walsh_2013}, 
suggesting that the formation of the cold disk is triggered by radiative cooling inside the Bondi radius ($\sim 120~\pc$).
The star formation rate for M87 is estimated as ${\rm SFR}\la 5\times 10^{-2}~\msunyr$ based on the $70\mu m$ infrared
luminosity \citep{Temi_2009, Calzetti_2010}.
The SFR would be considered as the upper limit because the observed aperture is considerably larger than the Bondi scales,
and there could be a level of contamination from synchrotron emission associated with the jet for M87.
This rate is consistent with our order-of-magnitude estimate of ${\rm SFR} \la M_\star /t_{\rm life} \simeq 5\times 10^{-2}~\msunyr$,
implying that a large fraction of the gas supplied from the Bondi radius is consumed by star formation, namely, 
${\rm SFR}/\dot{M}_{\rm B}\la 0.26$.
Presumably, a significant fraction of other giant ellipticals had experienced short episodes of star formation 
even after the galaxies have terminated major episodes of star formation at $z\la 1$ \citep[e.g.,][]{Thomas_2005,Perez-Gonzalez_2008}.
At lower redshifts, the number fraction of elliptical galaxies having ongoing nuclear star bursts, the so-called blue ellipticals, 
is as small as $\simeq 3~\%$ \citep{Tojeiro_2013}.
In addition, most nuclear disks in the centers of nearby ellipticals seem stable against star formation, i.e., $Q>1$ \citep{Boizelle_2017}.
Thus, the quiescent phases are expected to last the order of billions of years.

\subsection{Radiative luminosity vs. Bondi accretion rate}\label{sec:mdot-lumbol}

\begin{figure}
\begin{center}
\includegraphics[width=83mm]{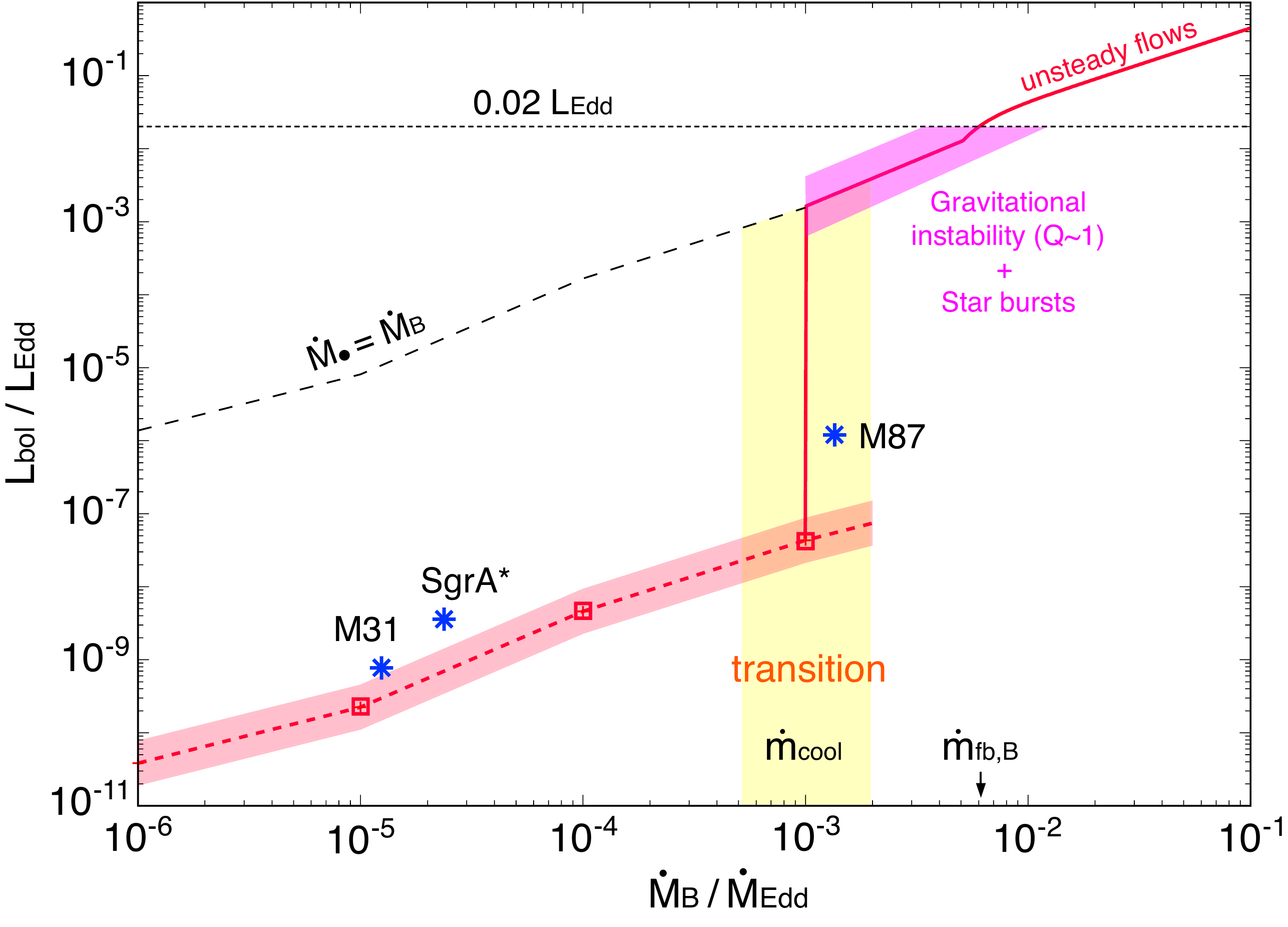}
\vspace{2mm}
\caption{Radiation luminosity ($10^{-10} \la L_{\rm bol}/L_{\rm Edd} \la 10^{-2}$) produced from accretion flows 
onto a BH at the Bondi rates of $10^{-6}\la \dot{M}_{\rm B}/\dot{M}_{\rm Edd} \la 10^{-2}$.
For lower values of $\dot{m}_{\rm B}\la \dot{m}_{\rm cool}(\sim 10^{-3})$, the radiation luminosity is significantly reduced from that with 
$\dot{M}_\bullet=\dot{M}_{\rm B}$ due to suppression of the BH feeding by convective motion (see Eq. \ref{eq:lbol_mb_low}).
The transition from radiatively inefficient accretion flows to cold accretion disks occurs at 
$4\times 10^{-4}\la \dot{M}_{\rm B}/\dot{M}_{\rm Edd} \la 2\times 10^{-3}$ (yellow shaded region).
After the transition, the accretion disk is likely to be unstable by its self-gravity, i.e., $Q\simeq 1$ (magenta shaded region; see Eq. \ref{eq:lbol_mb_high}).
At $L_{\rm bol}\ga 0.02~L_{\rm Edd}$ ($ \ga \dot{m}_{\rm fb,B}$), no steady accretion is realized due to strong BH feedback.
We also present observational results for Sgr A$^\star$ and BHs in M31 and M87 (blue asterisk).
}
\label{fig:summary_result}
\end{center}
\end{figure}

We here estimate the radiation luminosity ($L_{\rm bol}$) produced from accretion flows and BH feeding rate ($\dot{M}_\bullet$)
for a given Bondi accretion rate at a range of $10^{-6}\la \dot{M}_{\rm B}/\dot{M}_{\rm Edd} \la 10^{-2}$.
The results are summarized in Fig. \ref{fig:summary_result}.

Below the cooling threshold ($\dot{m}_{\rm B}<\dot{m}_{\rm cool}\sim 10^{-3}$), the accretion flows are adiabatic and 
the gas inflow rate decreases towards the center due to convection as shown in Fig. \ref{fig:2}.
In Paper I \citep{Inayoshi_2018}, we estimated the transition radius within which the inflow rate becomes constant 
because energy transport by thermal conduction would dominate that by convection.
The inferred accretion rate onto the central BH is 
\begin{align}
\frac{\dot{M}_\bullet}{\dot{M}_{\rm Edd}} & \simeq 1.5\times 10^{-6}~T_7^{-4/5}\nonumber\\
& \times \left(\frac{\alpha}{0.01}\right)^{0.37}
\left(\frac{\dot{m}_{\rm B}}{10^{-3}}\right)^{3/5}
\left(\frac{f_{\rm c}}{0.1}\right)^{2/5},
\end{align}
where $f_{\rm c}$ is the conductivity suppression factor because thermal conduction in the directions perpendicular
to magnetic fields can be suppressed. 
The value of the suppression factor has been discussed by various theoretical arguments and estimated as $f_{\rm c} \sim 0.1$ 
\citep[e.g.,][]{Narayan_Medvedev_2001,Maron_2004}.
For such low BH feeding rates, the radiative efficiency decreases with the accretion rate at $10^{-9} \la \dot{M}_\bullet/\dot{M}_{\rm Edd}\la 10^{-5}$
and has been estimated with radiation magneto-hydrodynamical simulations \citep[][$L\propto \dot{M}_\bullet^{1.7}$, see also
orange squares in Fig. \ref{fig:rad_eff}]{Ryan_2017}.
Therefore, adopting the radiation efficiency, the radiation luminosity model in this domain can be estimated as 
\begin{align}
\frac{L_{\rm bol}}{L_{\rm Edd}} & \simeq 3\times 10^{-8} \left(\frac{\alpha}{0.01}\right)^{0.63}
\left(\frac{f_{\rm c}}{0.1}\right)^{0.68}
\left(\frac{\dot{m}_{\rm B}}{10^{-3}}\right),
\label{eq:lbol_mb_low}
\end{align}
which is shown by the red shaded region in Fig. \ref{fig:summary_result}.
The width of the region reflects the uncertainties of the conductivity suppression factor $0.03\la f_{\rm c} \la 0.3$.
We note that if the BH feeding rate is equal to the Bondi accretion rate ($\dot{M}_\bullet=\dot{M}_{\rm B}$; black dashed), 
the radiation luminosities are much higher than that given by Eq. (\ref{eq:lbol_mb_low}).

For higher Bondi rates ($\dot{m}_{\rm B}\ga \dot{m}_{\rm cool}$), the accretion flow cools and forms a dense accretion disk
where the accretion rate is almost constant as a function of radius (see Fig. \ref{fig:2}).
Because the equation of state of the gas changes from adiabatic to isothermal one, the actual inflow rate becomes
comparable to the Bondi rate for isothermal gas and thus we obtain $\dot{M}_\bullet\simeq e^{3/2} \dot{M}_{\rm B}$.
Therefore, the radiation luminosity after the transition is estimated as
\begin{align}
\frac{L_{\rm bol}}{L_{\rm Edd}} & \simeq 1.6\times 10^{-3} 
\left(\frac{\dot{m}_{\rm B}}{10^{-3}}\right)^{1.27},
\label{eq:lbol_mb_high}
\end{align}
which would be valid for $\dot{m}_{\rm cool}\la \dot{m}_{\rm B}\la \dot{m}_{\rm fb,B}$, i.e.,
unless $L_{\rm bol}\ga 0.02~L_{\rm Edd}$ where radiative feedback suppresses gas inflows from larger radii.
This level of luminosities ($L_{\rm bol}/L_{\rm Edd}\simeq 10^{-3}-10^{-2}$) is consistent with the observational 
results for nearby Seyfert galaxies \citep{Koss_2017}.

For comparison, we present the observational results of Sgr A$^\star$ and BHs in M31 and M87 
(blue asterisks in Fig. \ref{fig:summary_result}; see also \citealt{Inayoshi_2018}). 
Those nearby SMBHs are known as quiescently accreting ones at rates of $\dot{M}_{\rm B}/\dot{M}_{\rm Edd} \sim 10^{-5}-10^{-3}$. 
The detailed gas properties in the nuclei are studied nicely by \textit{Chandra} X-ray observations resolving the angular sizes of their Bondi radii
\citep[e.g.,][and references therein]{Baganoff_2003,Garcia_2010,Russell_2015}.
Our theoretical estimate agrees well with the observational results for Sgr A$^\star$ and M31, and further explains the observed luminosity of M87,
which is several times higher than the adiabatic ones (red dashed) because of radiative cooling.

\subsection{Gas accretion from the circumnuclear disk to the Bondi scale}\label{sec:sfr}

So far, we have focused on the accretion physics around and inside the Bondi radius of the central BH, 
where a small centrifugal radius is set so that $R_{\rm C}<R_{\rm B}$.
We here briefly discuss gas supply to the Bondi scale from a circumnuclear disk with a size of $\sim10-100$~pc
\citep[CND;][]{TQM_2005, Ballantyne_2008}\footnote{This implicitly assumes a larger centrifugal radius. 
However, the gas inflow velocity could be a significant fraction of the Keplerian value in a CND
because of the Toomre instability and/or turbulence associated with star formation activities \citep{Wada_Norman_2002}.}.
Since CNDs would be massive reservoirs of molecular gas outside but very close to the Bondi radius
(e.g., \citealt{Hicks_2009} and \citealt{Izumi_2016} for warm and cold molecular gas, respectively), 
we naturally expect active star formation and/or gas feeding to the nuclear SMBH.
Interestingly, observations imply that the BH accretion rate would correlate with the SFR in the CNDs of Seyfert galaxies
\citep[e.g.,][]{Imanishi_2011,Diamond-Stanic_2012,Esquej_2014,Izumi_2016}.
This could suggest that star formation activities in CNDs (outside the Bondi scales) control the amount of gas fueling 
\citep[e.g.,][]{1989Natur.338...45S,Goodman_2003,TQM_2005,Tan_Blackman_2005,Vollmer_2008,Kawakatu_2008,Chamani_2017,Gan_2018}.

Numerical simulations studying the structure of a CND \citep{Wada_Norman_2002,Wada_2009} have found that 
SNe associated with active star formation lead to turbulence in a CND and thus the turbulent viscosity efficiently transports 
angular momentum outward and mass inward.
The turbulent velocity associated with energy injection by SNe is estimated as\footnote{
This equilibrium state would be realized unless super-bubbles created by SNe break out the disk scale height (cf. discussion in \S\ref{sec:QQQ}).
When the break-out timescale is longer than the sound crossing timescale in the disk, the disk would be highly turbulent
with typical velocity of $v_{\rm t}$.
}
\begin{equation}
v_{\rm t} \simeq \sqrt{ \frac{\eta E_{\rm SN}}{\tau_\star \Omega }},
\end{equation}
where $\eta$ is the efficiency per mass which denotes the fraction of the energy from SNe and $\tau_\star$ 
is the star formation timescale.
Presumably, the strength of an effective viscosity is as high as $\alpha_{\rm eff}\sim O(1)$ and thus 
the gas accretion rate through the disk is estimated as
\begin{align}
\dot{M}_{\rm d} 
\simeq 3\pi \alpha_{\rm eff} \eta E_{\rm SN}
\left( \frac{\Sigma}{\tau_\star}\right)
\left(\frac{r^3}{GM_\bullet}\right).
\end{align}
Assuming that the surface density follows a power law in the CND, $\Sigma =\Sigma_0 (r/r_{\rm out})^{-\gamma}$,
the CND mass is $M_{\rm d}=2\pi \Sigma_0r_{\rm out}^2[1-(r_{\rm in}/r_{\rm out})^{2-\gamma}]/(2-\gamma)$
and thus we obtain 
\begin{align}
\Sigma(r_{\rm in})=\frac{M_{\rm d}}{\pi r_{\rm out}^2} \cdot \frac{2-\gamma}{2}
\frac{x^{-\gamma}}{1-x^{2-\gamma}},
\end{align}
where $r_{\rm in(out)}$ is the inner (outer) radius of the CND and $x\equiv r_{\rm in}/r_{\rm out}$.
The gas accretion rate at the inner radius $r_{\rm in}$ is given by
\begin{align}
\dot{M}_{\rm d}(r_{\rm in}) &= \frac{3 \alpha_{\rm eff} \eta E_{\rm SN}M_{\rm d}}{\tau_\star}
\left(\frac{r_{\rm in}}{GM_\bullet}\right)
\frac{2-\gamma}{2}
\frac{x^{2-\gamma}}{1-x^{2-\gamma}},\nonumber\\
&\simeq  0.14~{\rm SFR}~\alpha_{\rm eff}M_8^{-1}  \left(\frac{r_{\rm in}}{2~\pc}\right) 
\sqrt{\frac{30~r_{\rm in}}{r_{\rm out}}}.
\end{align}
where ${\rm SFR}\equiv M_{\rm d}/\tau_\star$, $\eta = 10^{-3}~\msun^{-1}$, $E_{\rm SN}= 10^{51}~{\rm erg}$ and $\gamma=3/2$.
Adopting $r_{\rm in}\simeq R_{\rm B}$,
the gas accretion rate through the Bondi radius is given by 
\begin{align}
\dot{M}_{\rm d}(R_{\rm B}) &\simeq  0.14~{\rm SFR}~\alpha_{\rm eff}T_7^{-1} \sqrt{\frac{30~r_{\rm in}}{r_{\rm out}}}.
\end{align}
Therefore, star formation activities in the nuclear region in a CND at a rate of ${\rm SFR}\simeq 0.1~\msunyr$ 
provide an inflow rate at $\sim 10^{-2}~\msunyr$ within the Bondi radius of an SMBH.
For the typical mass of BHs hosted in Seyfert galaxies/AGN ($M_\bullet \sim 10^7- 10^8~\msun$), the Bondi accretion rate is 
high enough to form a cold, dense accretion disk, leading to a high BH feeding rate of $\dot{M}_\bullet \simeq \dot{M}_{\rm B}$
and a high luminosity of $L_{\rm bol}/L_{\rm Edd}\simeq 10^{-3}-10^{-2}$ \citep[see also][]{Koss_2017}. 
The result is consistent with the fact that the BH feeding rate inferred from the nuclear luminosity tightly correlates with 
the SFR in the CND ($<100~\pc$), i.e., $\dot{M}_\bullet /{\rm SFR} \sim O(0.1)$ \citep[e.g.,][]{Diamond-Stanic_2012,Esquej_2014,Izumi_2016}.
On the other hand, for higher BH masses ($M_\bullet \gg 10^8~\msun$), the Bondi rate would not be high enough to 
trigger the cooling transition.

\subsection{Caveats}
\subsubsection{magnetic fields}

In this work, we do not take into account MHD effects explicitly, but adopt the $\alpha$-viscosity prescription 
to treat angular momentum transport, mimicking some properties of MRI turbulence.
The properties of radiatively inefficient MHD accretion flows are affected qualitatively and quantitatively by the initial configuration 
and boundary conditions for magnetic fields \citep[e.g.,][]{Stone_Pringle_2001,INA_2003,McKinney_2012,Narayan_2012,Yuan_2012b}.
Assuming that a toroidal magnetic field or multiple poloidal loops is initially set or injected with accreting matter from the outer boundary,
the accretion flow is dominated by turbulent motion driven by MRI and heating associated with magnetic reconnection.
Through the dissipation process, the strength of the magnetic field is saturated at some levels where the ratio of the gas pressure to 
the magnetic pressure  is $\beta \equiv P_{\rm gas}/P_{\rm m} \simeq 10^2-10^3$.
This type of accretion flow is qualitatively similar to the results obtained from hydrodynamical simulations adopting $\alpha$-viscosity 
\citep[e.g.,][]{Inayoshi_2018}.

On the other hand, when a magnetic field with poloidal topology is dragged inward by accreting matter and accumulated at the vicinity of the BH,
the accretion flow behaves in a very different way.
In fact, the flow is less turbulent but is likely to produce outflows and/or jets associated with the amplified strong magnetic field.
\cite{Yuan_2012b} studied the detailed properties of this type of MHD flow and concluded that the flow is more related to
the so-called adiabatic inflow-outflow solution \citep[][]{Blandford_Begelman_2004}, 
where matter can accrete at very low rates due to strong outflows which carry energy and angular momentum away.

It is also worthy investigating this issue for gas accretion from outside the Bondi radius where 
the geometry and binding energy of the injected mass are different from those studied in previous work.
In this case, a more realistic situation is that magnetic fields at larger scales fluctuate with time due to 
gas turbulent motion, and then are dragged inward by accreting matter.
We leave this to future work (but note that \cite{Igumenshchev_Narayan_2002} briefly discussed the issue for spherical accretion).

\subsubsection{mechanical feedback due to winds}

Mechanical feedback due to AGN winds is expected to play an important role on the evolution of galaxies 
when the BH feeding rate is sufficiently low 
\citep[e.g.,][]{Murray_2005,Ostriker_2010,Gaspari_2012,Choi_2015}.
Assuming a phenomenological model \citep{Ostriker_2010}, one can estimate an upper bound 
on the momentum input by winds as
\begin{align}
\dot{P}_{\rm w} &=\dot{M}_{\rm w}v_{\rm w} = \dot{M}_{\rm inf}v_{\rm w}\cdot \frac{\eta}{1+\eta} \\
& \simeq 3\times 10^{-4}~\frac{L_{\rm Edd}}{c} \left(\frac{\dot{M}_{\rm inf}}{10^{-3}~\dot{M}_{\rm Edd}}\right)
\left(\frac{v_{\rm w}}{10^{4}~{\rm km~s^{-1}}}\right),\nonumber
\end{align}
where $\dot{M}_{\rm w}$ is the wind mass-loss rate, $v_{\rm w}$ is the wind velocity,
$\dot{M}_{\rm inf}$ is the mass inflow rate at the inner grid of $r=r_{\rm in}$,
$\eta$ is the ratio of the wind mass-loss rate to the BH feeding rate ($\dot{M}_\bullet \equiv \dot{M}_{\rm inf}-\dot{M}_{\rm w}$),
and $\dot{M}_\bullet $ is assumed to be 10\% of the inflow rate, i.e., $\eta =9$.
The outflowing material pushes the accreting gas bound by the BH gravity and could form an momentum-driven expanding shell. 
The strength of gravitational force exerting the shell with a mass of $M_{\rm sh}$ at a radius of $r$ is estimated as
\begin{equation}
|F_{\rm grav}| = \frac{GM_\bullet M_{\rm sh}}{r^2} \simeq \frac{L_{\rm Edd}}{c} \frac{1}{4\pi r^2}\int  \rho \kappa_{\rm es} dV,
\end{equation}
where mass accretion to the shell from larger scales is neglected.

Let us compare the strength of momentum input and gravitational force on a mass shell for the two 
different accretion regimes.
For adiabatic cases with $\dot{m}_{\rm B}< 10^{-3}$, where the mass inflow rate is reduced from the Bondi rate due to turbulent motion as 
$\dot{M}_{\rm inf}/\dot{M}_{\rm B} \la 10^{-2}$, the strength of mechanical feedback is estimated as 
$\dot{P}_{\rm w} \la 3\times 10^{-3}~\dot{m}_{\rm B}L_{\rm Edd}/c$.
Using the density profile of a CDAF near the mid-plane (Eq. 21 in \citealt{Inayoshi_2018}), one can estimate the gravitational force 
on the shell as $|F_{\rm grav}| \sim 6\times 10^{-2}~\dot{m}_{\rm B} (L_{\rm Edd}/c) (r/R_{\rm C})^{1/2}$ and thus 
$\dot{P}_{\rm w}/|F_{\rm grav}| \la 0.05$.
Next, for a higher rate of $\dot{m}_{\rm B} \sim 10^{-3}$, where the inflowing rate through a cold disk is comparable to the Bondi rate
without suppression of accretion, the level of momentum input is boosted up to $\dot{P}_{\rm w} \sim 3\times 10^{-1}~\dot{m}_{\rm B}L_{\rm Edd}/c$,
but is not strong enough to blow the dense accretion disk away (i.e., $\dot{P}_{\rm w}/|F_{\rm grav}| \ll 1$).
Therefore, even in this case where a generous fraction of the inflowing gas is ejected in a high velocity, 
mechanical feedback does not affect the gas dynamics both in CDAF and cold disk accretion cases.

On the other hand, mechanical feedback due to winds would be likely to blow the gas away {\it near the poles} where the gas density is sufficiently low
and weak outflows are produced even without explicitly assuming AGN winds (see Fig. \ref{fig:cont}). 
As a natural outcome, biconical outflows with opening angles of $\sim 30^\circ~(70^\circ)$ would be produced from
geometrically thick (thin) disk in lower (higher) accretion-rate regimes.
To explore the nature and origin of outflows both for CDAF and cold disk accretion solutions is left for future investigations
(but note that \citealt{Proga_2000} discussed wind production from the nuclear disk at an accretion rate of $\sim \dot{M}_{\rm Edd}$).

\section{Summary and conclusion}
\label{sec:sum}
We study the properties of rotating accretion flows onto SMBHs using
axisymmetric two-dimensional radiation hydrodynamical simulations with radiative cooling and BH feedback.
The simulations resolve the accretion dynamics of gas outside the BH influence radius through an inner accretion disk.
For lower Bondi accretion rates at $\dot{M}_{\rm B}\ll 10^{-3}~\dot{M}_{\rm Edd}$,
the BH feeding is suppressed by several orders of magnitude from the Bondi rate due to turbulent motion
with outflows to the Bondi radius nearly balancing inflows.
Thus, the radiative luminosity results in as low as $\sim 10^{-10}-10^{-7}~L_{\rm Edd}$. 
For higher rates of $\dot{M}_{\rm B}\ga 10^{-3}~\dot{M}_{\rm Edd}$, the optically-thin accreting gas cools via free-free emission 
and forms a geometrically-thin disk, which feeds the BH efficiently and increases the radiative luminosity to $\ga 10^{-3}~L_{\rm Edd}$.

The transitional behavior of accreting BHs in galactic nuclei from the radiative inefficient phases to cold disk accretion naturally explains 
(1) the reason for the offset between the observed luminosities and the simplest theoretical predictions for nearby quiescent SMBHs, and 
(2) the conditions to fuel gas into the nuclear SMBH.
In addition, the cold disks formed in galactic nuclei tend to be gravitationally unstable and lead to star formation
when the Bondi rate is as high as $ \dot{M}_{\rm B} \ga 10^{-2}~\msunyr$.
This plausibly explains the correlation between star formation rates and BH feeding rates in Seyfert galaxies.

\section*{Acknowledgements}
We thank Feng Yuan, Luis Ho, Takuma Izumi, Defu Bu and Zhaoming Gan for useful discussions.
This work is partially supported by the National Key R\&D Program of China (2016YFA0400702), 
and the National Science Foundation of China (11721303), the Simons Foundation through 
the Simons Society of Fellows (Kohei Inayoshi), and supported by Program for Establishing a Consortium
for the Development of Human Resources in Science and Technology, Japan Science and Technology Agency and 
Japan Society for the Promotion of Science (JSPS) KAKENHI (18K13584; Kohei Ichikawa). 
RK acknowledges financial support via the Emmy Noether Research Group on Accretion Flows 
and Feedback in Realistic Models of Massive Star Formation funded by the German Research Foundation 
(DFG) under grant no. KU 2849/3-1 and KU 2849/3-2.
Numerical computations were carried out on Cray XC50 at the Center for Computational Astrophysics 
of the National Astronomical Observatory of Japan.

\bibliographystyle{mnras}
{\small
\bibliography{ref}
}
\appendix

\section{Effective viscosity by the global instability in a gravitationally unstabel disk}
\label{sec:app1}

After the transition to a geometrically-thin, cold and dense accretion disk, 
the disk is likely to be unstable to its self-gravity, where the Toomre parameter becomes close to unity (i.e., $Q\la 1$).
In order to capture the Toomre instability, we adopt a toy model for the effective viscosity in a gravitationally unstable disk as
\begin{equation}
\alpha_{\rm eff} = \alpha_0 + \alpha_{\rm max}\exp(-Q^A/B),
\end{equation}
In spite of uncertainties of those parameters, \cite{Takahashi_2013} found that for a wide range of $(A,B)$
this model for the effective viscosity can reproduce the results of three-dimensional hydrodynamical 
simulations of a circumstellar disk under continuous gas supply from larger scales \citep{Machida_2010}.

\begin{figure}
\begin{center}
\includegraphics[width=80mm]{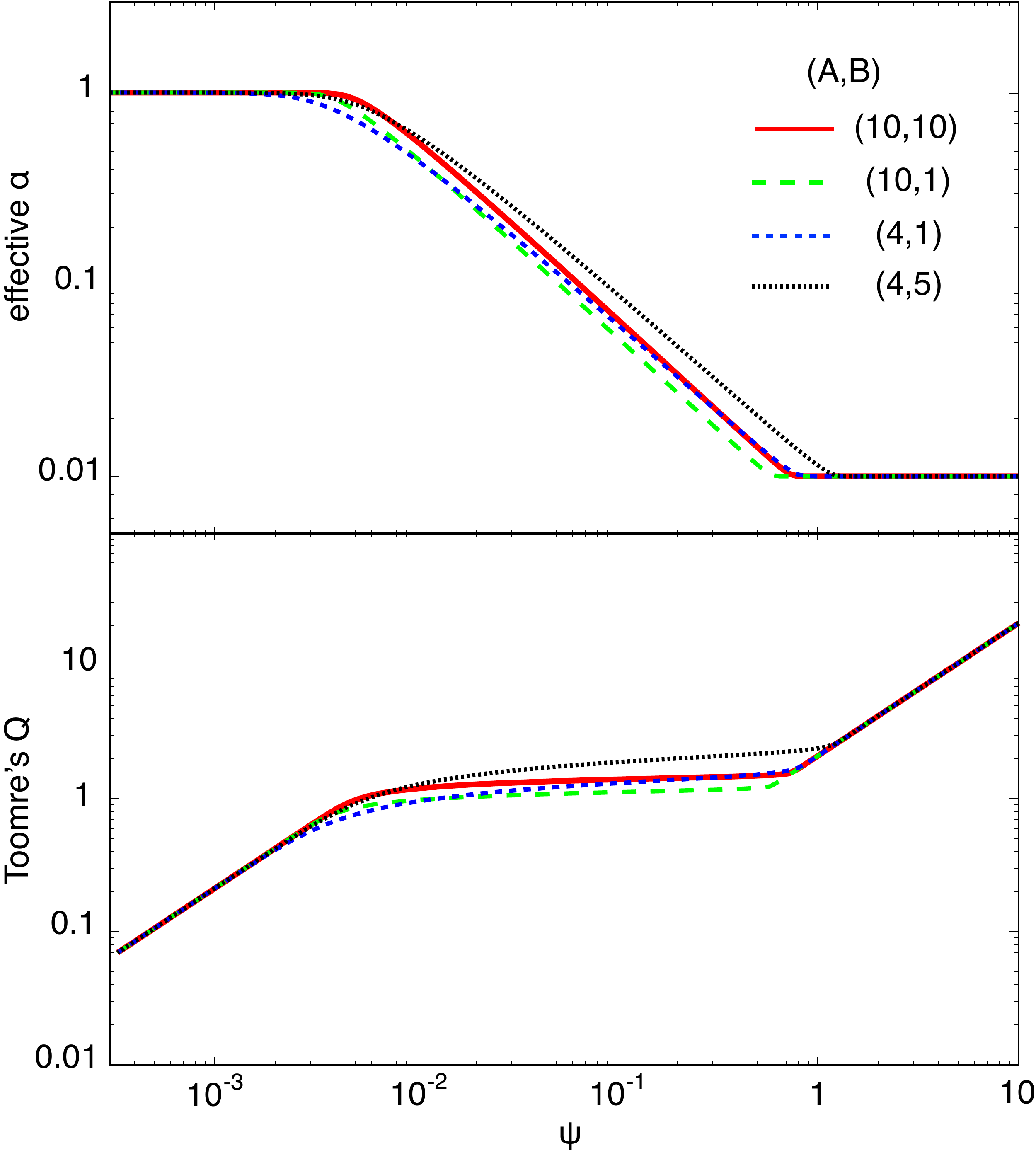}
\caption{
Dependence of the effective viscous parameter $\alpha_{\rm eff}$ and Toomre parameter $Q$ on
the choice of the free model parameters.
The results hardly depend on ($A,B$) but on $\psi \equiv T_{\rm min,4}^{3/2}M_8^{-1}(\dot{m}_{\rm B}/10^{-3})^{-1}$.
}
\label{fig:alpha_Q_eff}
\end{center}
\end{figure}

In Fig. \ref{fig:alpha_Q_eff}, we show the effective viscosity (top) and Toomre parameter (bottom)
as a function of
\begin{equation}
\psi \equiv T_{\rm min,4}^{3/2}M_8^{-1}\left(\frac{\dot{m}_{\rm B}}{10^{-3}}\right)^{-1}
\end{equation}
for different combinations of $(A,B)$.
For $\psi\ga 1$, the Toomre parameter is significantly above unity and thus $\alpha_{\rm eff}\simeq \alpha_0(=0.01)$.
With $\psi$ decreasing ($10^{-2}< \psi < 1$), the Toomre parameter is almost constant at $Q\simeq 1$ and 
the effective viscosity increases.
At the intermediate region, gravitational torques could transport mass and angular momentum so that the disk structure 
results in a marginally stable state.
For $\psi \ll 10^{-2}$, the effective viscosity is saturated at $\alpha_{\rm eff}=1$ and the Toomre parameter becomes 
substantially lower than unity.
Since the disk no longer keeps stable accretion, therefore the disk would be likely to fragment into clumps.
We set the two free parameters to $(A,B)=(10,1)$ in \S\ref{sec:QQQ}.

\end{document}